\documentclass[11pt,letterpaper]{article}

\usepackage{amsmath, amssymb}
\usepackage{graphicx}
\usepackage{pstricks}
\usepackage{pst-plot}
\usepackage{pst-math}

\title{The illusion of acceleration in the retarded Li\'{e}nard-Wiechert electromagnetic field}
\author{C\u{a}lin Galeriu \\
Military Technical Academy \lq\lq Ferdinand I\rq\rq \\
Bvd. George Co\c{s}buc nr. 39-49, Bucharest, Romania \\
{\it calin.galeriu@mta.ro}}
\date{}

\begin{document}

\maketitle

\begin{abstract}
{\it It is generally assumed that 
the retarded Li\'{e}nard-Wiechert electromagnetic field produced by a point particle 
depends on the acceleration of that source particle. 
This dependence is not real, it is an illusion.
The true electromagnetic interaction is time symmetric (half retarded and half advanced) and 
depends only on the positions and velocities of the electrically charged particles. 
A different acceleration of the retarded source particle will result in a different position and 
velocity of the advanced source particle, changing in this way the Lorentz force felt by the 
test particle.}
\end{abstract}

\section{Introduction}

While the algebraic formalism of classical electrodynamics is well established,
we would also like to have a comprehensive geometrical derivation of the 
electrodynamic forces between 
electrically charged point particles. From a geometrical point of view,
the interacting particles are represented by worldlines in Minkowski space, and 
we hope that, with the proper theoretical structure in place, 
\lq\lq the laws of physics can find their most complete
expression as interrelations between these worldlines\rq\rq \cite{minkowski}.

We develop our theory by relying on geometrical intuition,
as well as on physical and philosophical arguments.
We start with the simplest case of electrostatic interaction, then we allow for 
constant velocities, then we allow for constant accelerations, 
and finally we allow for variable accelerations.
Along the way we introduce new postulates, expanding the theoretical framework,
until we end up with a time symmetric action-at-a-distance theory
that reproduces the classical electrodynamics theory as a second order approximation.

\section{Source particle at rest}

Consider a source particle at rest, with electric charge $Q$, and a field point at $\vec{R}$,
where $\vec{R}$ is the displacement vector from the source particle to the field point.
The particle is the source of an electric field $\vec{E}$ that points in the radial direction of $\vec{R}$
and has the magnitude of $Q/R^2$ (in Gaussian units), where $R$ is the distance between the source
particle and the field point. In this reference frame the magnetic field $\vec{B}$
is zero and the Lorentz force has only an electric part, that is $\vec{F} = q \, \vec{E}$.

A Lorentz transformation could bring us to a reference frame
where the source particle is moving with constant velocity, the magnetic field
is not zero, and the Lorentz force, for a test particle in motion, has also a magnetic part. 
This Lorentz transformation in effect introduces the velocity dependent magnetic interaction as a relativistic efect.
However, even in this case, the electromagnetic interaction is due to the same four-force as before, 
which has different spatial and temporal components in different inertial reference frames \cite{galeriu3D}.
For this reason in Section 2 we only investigate the case of a source particle at rest. 

\subsection{Test particle at rest}

Consider a test particle at rest, with electric charge $q$.

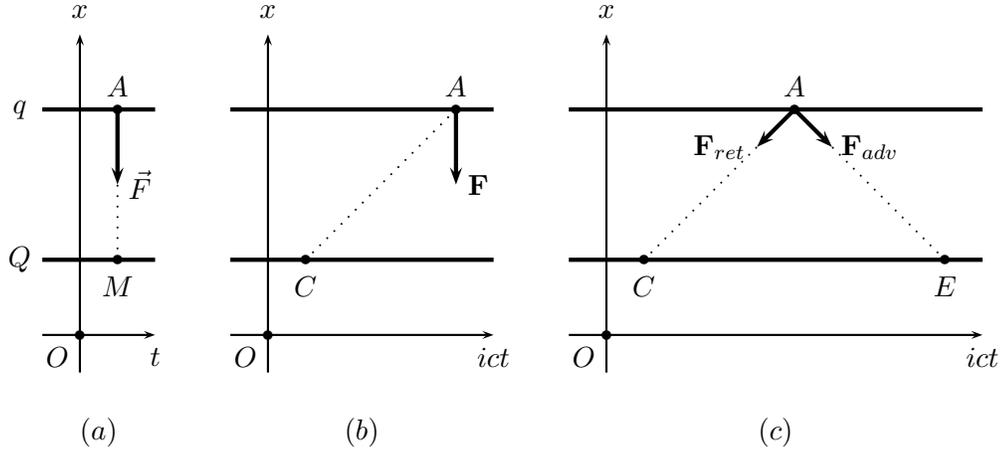
\begin{figure}[h!]
\begin{center}
\begin{pspicture}(-0.5,0)(13,6)
\rput(-0.3,2.5){$Q$}
\rput(-0.3,4.5){$q$}
\psline{->}(0,1.5)(1.5,1.5)
\psline{->}(0.5,1)(0.5,5.5)
\psdot(0.5,1.5)
\rput(0.2,1.2){$O$}
\rput(0.5,5.8){$x$}
\rput(1.5,1.2){$t$}
\psline[linewidth=1.6pt](0,2.5)(1.5,2.5)
\psline[linewidth=1.6pt](0,4.5)(1.5,4.5)
\psdot(1,2.5)
\rput(1,2.15){$M$}
\psdot(1,4.5)
\rput(1,4.8){$A$}
\psline[linewidth=1.6pt]{->}(1,4.5)(1,3.5)
\rput(1.3,3.5){$\vec{F}$}
\psline[linestyle=dotted](1,3.5)(1,2.5)
\rput(0.75,0.2){$(a)$}
\psline{->}(2.5,1.5)(6,1.5)
\psline{->}(3,1)(3,5.5)
\psdot(3,1.5)
\rput(2.7,1.2){$O$}
\rput(3,5.8){$x$}
\rput(6,1.2){$ict$}
\psline[linewidth=1.6pt](2.5,2.5)(6,2.5)
\psline[linewidth=1.6pt](2.5,4.5)(6,4.5)
\psdot(3.5,2.5)
\rput(3.5,2.15){$C$}
\psdot(5.5,4.5)
\rput(5.5,4.8){$A$}
\psline[linewidth=1.6pt]{->}(5.5,4.5)(5.5,3.5)
\rput(5.8,3.5){$\mathbf{F}$}
\psline[linestyle=dotted](5.5,4.5)(3.5,2.5)
\rput(4.25,0.2){$(b)$}
\psline{->}(7,1.5)(12.5,1.5)
\psline{->}(7.5,1)(7.5,5.5)
\psdot(7.5,1.5)
\rput(7.2,1.2){$O$}
\rput(7.5,5.8){$x$}
\rput(12.5,1.2){$ict$}
\psline[linewidth=1.6pt](7,2.5)(12.5,2.5)
\psline[linewidth=1.6pt](7,4.5)(12.5,4.5)
\psdot(8,2.5)
\rput(8,2.15){$C$}
\psdot(10,4.5)
\rput(10,4.8){$A$}
\psdot(12,2.5)
\rput(12,2.15){$E$}
\psline[linewidth=1.6pt]{->}(10,4.5)(9.5,4)
\rput(9,4){$\mathbf{F}_{ret}$}
\psline[linestyle=dotted](9.5,4)(8,2.5)
\psline[linewidth=1.6pt]{->}(10,4.5)(10.5,4)
\rput(11,4){$\mathbf{F}_{adv}$}
\psline[linestyle=dotted](10.5,4)(12,2.5)
\rput(9.75,0.2){$(c)$}
\end{pspicture}
\caption{Examples of (a) instantaneous, (b) retarded, and (c) time symmetric
electrostatic interactions.}
\label{fig:fig1}
\end{center}
\end{figure}

When the interaction is instantaneous, as in classical Newtonian physics, 
the direction of the force $\vec{F}$ is along the straight line connecting the two particles.
Let this also be the direction of the $x$ axis of our reference frame.
Without loss of generality we assume that both particles have a
positive $x$ coordinate, with the test particle being farther away from the origin.
We also assume that $Q > 0$ and $q < 0$, 
such that the electrostatic force is attractive.
This situation is represented graphically in Figure \ref{fig:fig1} (a), where the test particle
is at $A$, the source particle is at $M$, and the segment $AM$ has length $R$.

When the interaction is retarded, 
and the 3D Euclidean space is embedded into the 4D Minkowski space,
the direction of the four-force 
$\mathbf{F} = (\vec{F}, 0)$ 
no longer matches the direction of the straight line connecting the two particles.
This situation is represented graphically in Figure \ref{fig:fig1} (b), where the test particle
is at $A$, the retarded source particle is at $C$,
and the segment $AC$ has null length. 
The mismatch between the two directions is very puzzling. 

When the interaction is time symmetric,
the retarded and the advanced four-forces could point towards their corresponding source particles,
and thus have the correct intuitive direction.
By postulate (Postulate I) we assume that this indeed happens \cite{galeriuAAAD}.
As seen from Figure \ref{fig:fig1} (c),
due to the symmetry of this configuration under a time inversion operation,
the retarded $\mathbf{F}_{ret}$ and 
the advanced $\mathbf{F}_{adv}$ four-forces must be the mirror image of each other.
\begin{eqnarray}
\mathbf{F}_{ret} = \Big(\frac{Q q}{2 R^2}, 0, 0, i \frac{Q q}{2 R^2}\Big) \label{equation1} \\
\mathbf{F}_{adv} = \Big(\frac{Q q}{2 R^2}, 0, 0, - i \frac{Q q}{2 R^2}\Big) \label{equation2}
\end{eqnarray}
The spatial (real) components add together, 
but the temporal (imaginary) components
cancel each other.
The resultant electrostatic four-force
\begin{equation}
\mathbf{F} = \mathbf{F}_{ret} + \mathbf{F}_{adv} = \Big(\frac{Q q}{R^2}, 0, 0, 0\Big)
\label{equation3}
\end{equation}
matches the standard expression. The total four-force $\mathbf{F}$ is orthogonal to the
four-velocity of the test particle, even though the individual retarded and advanced 
four-forces are not.
This situation is represented graphically in Figure \ref{fig:fig1} (c), where the test particle
is at $A$, the retarded source particle is at $C$, and the advanced source particle is at $E$.

We also notice that, in this very simple case when both particles are at rest, 
the retarded four-force with which the particle at $C$ acts on the particle at $A$
is equal in magnitude but opposite in direction to
the advanced four-force with which the particle at $A$ acts on the particle at $C$.
The same holds true for the interaction between the particles at $A$ and at $E$.

Several other authors have also considered this type of interaction, where
the four-forces (the exchanges in four-momentum) are directed along the
null segments connecting the interacting particles.
Gen Yoneda \cite{GenYoneda} states that \lq\lq it is natural to assume\rq\rq \ that
the retarded and advanced four-forces have the same direction as the displacement four-vectors connecting the 
interacting particles, and names this \lq\lq the parallel condition\rq\rq. 
Fokker \cite{Fokker1965} postulates that
\lq\lq the mass vector of the transferred
fragment is directed along the zero interval, so that the mass of the
fragment is equal to zero.\rq\rq \ 
Synge \cite{synge} is even more explicit: \lq\lq At the event $P$ some entity possessing momentum and energy
(i.e. a momentum-energy 4-vector) leaves the first particle and travels in a straight line with the
fundamental velocity $c$ (i.e. its world line is a straight null line) to meet the other particle at the
event Q, where it is absorbed.\rq\rq \ 
This entity, possessing energy $E$ and linear momentum $E/c$, is like a real photon, with one difference: 
the energy $E$ is positive for a repulsive force, but negative for an attractive force. 
This classical counterpart of a quantum mechanical exchange particle
is called \lq\lq telehapsis\rq\rq \ by Fokker and \lq\lq internal impulse\rq\rq \ by Synge.
During their interaction in spacetime 
the two particles are exchanging a continuous stream of internal impulses,
and it is quite remarkable the fact that
\lq\lq for particles with constant proper
masses and relative velocity small compared with c,
the inverse square law is a consequence of the mechanical
laws of conservation\rq\rq \cite{Synge1935}.
Due to the conservation of total four-momentum whenever the internal impulses are emitted or absorbed,
this interaction mechanism \lq\lq embodies the simplest available generalization to relativity of Newton's law of 
action and reaction\rq\rq \cite{synge}.

As noticed by Synge, this model of interaction must be time symmetric.
Since only the emission (or only the absorption) of a single real photon cannot leave the rest mass of the 
test particle invariant (due to the triangle inequality in Minkowski space, applied to four-momenta),
in order for this rest mass to remain constant
we need to have a simultaneous exchange of internal impulses 
with both the retarded and the advanced source particles. 

The assumed time symmetry of the interaction has important physical and philosophical implications.
For those investigating the nature and ontology of spacetime, as emphasized  by Vesselin Petkov,
\lq\lq the main question is whether the world is three-dimensional or four-dimensional\rq\rq \cite{VesselinPetkov}.
Only the four-dimensionalist view is compatible with time symmetric interactions.
Once the growing block universe theory is rejected and eternalism is embraced, we conclude that
\lq\lq unlike the pre-relativistic division of events, the relativistic division does not affect the existence of the events -- the events
in the past light cone, the event $O$, and the events in the future light cone are all {\it equally} existent\rq\rq \cite{VesselinPetkov}
and that
\lq\lq if we consider the worldline of a particle in spacetime, the worldline is a monolithic four-dimensional entity 
which exists timelessly in the frozen world of Minkowski spacetime\rq\rq \cite{VesselinPetkov}.
From the point of view of {\it classical physics} the future is already there and cannot be changed. 
Only by introducing {\it quantum mechanics} into the theory one can hope to address complex topics such as
the flow of time, the human conscience, the animal conscience, free will, and acausal change.
 
\subsection{Test particle with only radial velocity $\vec{v} = (v_x, 0, 0)$}

When the test particle has a purely radial velocity $\vec{v} = (v_x, 0, 0)$, 
that means when the velocity has the same radial direction as the displacement vector $\vec{R} = (R, 0, 0)$, 
the electric field $\vec{E} = (Q / R^2, 0, 0)$, and the electrostatic force $\vec{F} = (Q q / R^2, 0, 0)$,
the four-force becomes 
\begin{equation}
\mathbf{F} = \Big(\gamma \vec{F}, i \frac{\gamma}{c} \vec{F} \cdot \vec{v} \Big)
= \Big(\gamma \frac{Q q}{R^2}, 0, 0, i \gamma \frac{v_x}{c} \frac{Q q}{R^2}\Big),
\label{eq:fourforce}
\end{equation}
where $\gamma = 1 / \sqrt{1 - v^2 / c^2}$
is the Lorentz factor and $v = |v_x|$ is the speed.

We notice that the four-force depends explicitly on the velocity of the test particle. 
\lq\lq But, from a {\it geometrical} point of view, a point in Minkowski space is just a fixed point
-- it does not have a velocity!\rq\rq \cite{galeriuAAAD}
The four-force, through the distance $R$, also depends implicitly on the velocity of the source particle,
since this distance is measured in the inertial reference frame where the source particle is instantaneously at rest.
If given only two spacetime points connected by a segment of null length, it is impossible to find
a relativistically invariant non-zero distance. 
How can we explain this dependence of the four-force on the velocities of the interacting particles? 
We have to acknowleddge the fact that the material point particle model,
perfectly valid in the 3D Euclidean space, looses its applicability in the 4D Minkowski space.
Here, in spacetime, the interaction takes place not between points, but between
segments of infinitesimal length along the worldlines of the particles. Elementary particles
cannot have a null dimension along the time axis of their proper reference frame.
While exploring the concept of inertia, Kevin Brown has also reached the conclusion that
\lq\lq even an object with zero spatial extent has non-zero temporal extent\rq\rq \cite{brown}.

A theory of time symmetric action-at-a-distance
electrodynamic interactions between length elements along the worldlines of the particles
has been proposed by Fokker \cite{fokker} as early as 1929.
The corresponding effective segments of infinitesimal length (\lq\lq entsprechenden effektiven Elementen\rq\rq), 
who have the important property that their
end points are connected by light signals, enter the expression of the Fokker action
in a  totally symmetric manner, with no distinction between source and test particles. 
This leads to the conservation of four-momentum, most clearly stated by Wheeler and Feynman:
\lq\lq The impulse communicated to $a$ over the portion $d\alpha$ of its world line via retarded forces,
for example, from the stretch $d\beta$ of the world line of $b$ is equal in magnitude and opposite in sign 
to the impulse transfer from $a$ to $b$ via advanced forces over the same world line intervals
({\it equality of action and reaction}).\rq\rq \ \cite{Wheeler1949}
Their algebraic reformulation of Fokker's theory, making use of Dirac delta functions, 
was demonstrated to be equivalent to classical electrodynamics. 
Unfortunately, since Wheeler and Feynman have assumed that the Dirac delta function is non-zero in just one point,
the geometrical insight that we are looking for was lost.
In order to safeguard our geometrical intuition, 
we assume instead that the Dirac delta function is non-zero inside an infinitesimal interval
\cite{Amaku2021RBEF, galeriu2023}.
Another important observation is that, unlike in our theory,
Fokker \cite{fokker} and Wheeler and Feynman \cite{Wheeler1949} 
do not implement \lq\lq the parallel condition\rq\rq, 
but keep each of the individual retarded and advanced four-forces
orthogonal to the four-velocity of the particle on which they act uppon. 

Surprisingly, even Hermann Minkowski was thinking about this alternative way of describing the interaction
between two particles, 
seen as worldline segments of infinitesimal length whose endpoints are connected by light signals,
as demonstrated by the words that he wrote when presenting his formula for the gravitational four-force.
\lq\lq Imagine the spacetime threads of $F$ and $F^*$ with the main lines in them.
Let us take an infinitely small element $BC$ on the main line of $F$, 
further on the main line of $F^*$,
$B^*$ is the point light source of $B$  
and $C^*$ is the point light source of $C$;\rq\rq \cite{minkowski}

Another way to proclaim that the interaction does not take place between points in Minkowski space, 
but between corresponding length elements,
is to reffer to the infinitesimal but non-zero \lq\lq thickness of the light cone\rq\rq. 

Hugo Tetrode \cite{tetrode} mentions that only the infinitesimal neighborhood of
the lightcone (\lq\lq da\ss \ nur die infinitesimale Umgebung des Lichtkegels\rq\rq) with vertex at the field
point contributes to the electromagnetic potentials.

Nicholas Wheeler \cite{NicholasWheeler} draws a lightcone 
with \lq\lq some small but finite\rq\rq\ thickness and writes:
\lq\lq If the lightcone had \lq thickness\rq \ then the presence
of the Doppler factor in (456) could be understood qualitatively to
result from the relatively \lq longer look\rq \ that the field point gets at
approaching charges, the relatively \lq briefer look\rq \ at receding charges.\rq\rq

Kevin Brown \cite{brown} also draws a diagram in which
\lq\lq the light cone is shown with a non-zero thickness 
to illustrate that the duration of time spent by each particle as it passes through the light cone 
depends on the speed of the particle.\rq\rq 

\begin{figure}[h!]
\begin{center}
\begin{pspicture}(-5.5,-1.5)(5.5,2)
\pspolygon*[linestyle=none,linecolor=yellow](-2,2)(-4,0)(-2.5,-1.5)(-1.5,-1.5)(-3,0)(-1,2)
\psline{->}(-5.5,-1)(-0.5,-1)
\psline{->}(-5,-1.5)(-5,1.5)
\rput(-0.5,-1.3){$i c t$}
\rput(-5,1.8){$x \ y \ z$}
\psdot(-5,-1)
\rput(-5.3,-1.3){$O$}
\psline(-4.5,0)(-0.5,0)
\psline[showpoints=true,linewidth=1.5pt](-4,0)(-3,0)
\rput(-4,-0.3){$C$}
\rput(-3,-0.3){$D$}
\psline[linestyle=dotted](-4,0)(-2,2)
\psline[linestyle=dotted](-3,0)(-1,2)
\psline[linestyle=dotted](-4,0)(-2.5,-1.5)
\psline[linestyle=dotted](-3,0)(-1.5,-1.5)
\psline(-4,1)(-1,1)
\psline[showpoints=true,linewidth=1.5pt](-3,1)(-2,1)
\rput(-3,1.3){$A$}
\rput(-2,1.3){$B$}
\pspolygon*[linestyle=none,linecolor=yellow](2,2)(4,0)(2.5,-1.5)(3.5,-1.5)(5,0)(3,2)
\psline{->}(0.5,-1)(5.5,-1)
\psline{->}(1,-1.5)(1,1.5)
\rput(5.5,-1.3){$i c t$}
\rput(1,1.8){$x \ y \ z$}
\psdot(1,-1)
\rput(0.7,-1.3){$O$}
\psline(1.5,0)(5.5,0)
\psline[showpoints=true,linewidth=1.5pt](4,0)(5,0)
\rput(4,-0.3){$E$}
\rput(5,-0.3){$F$}
\psline[linestyle=dotted](4,0)(2,2)
\psline[linestyle=dotted](5,0)(3,2)
\psline[linestyle=dotted](4,0)(2.5,-1.5)
\psline[linestyle=dotted](5,0)(3.5,-1.5)
\psline(2,1)(5,1)
\psline[showpoints=true,linewidth=1.5pt](3,1)(4,1)
\rput(3,1.3){$A$}
\rput(4,1.3){$B$}
\end{pspicture}
\caption{A test particle $AB$, at rest relative to the source particle, 
feels a retarded interaction from $CD$ and an advanced interaction from $EF$.}
\label{fig:fig3}
\end{center}
\end{figure}
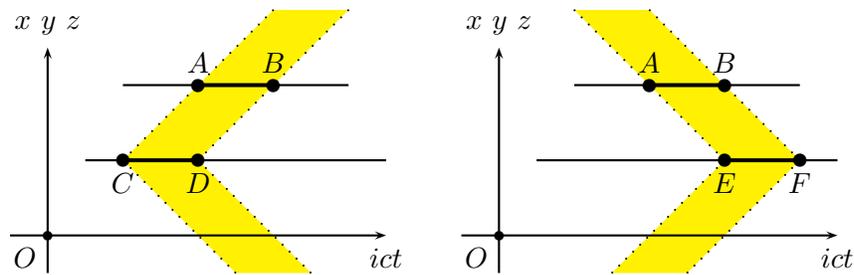

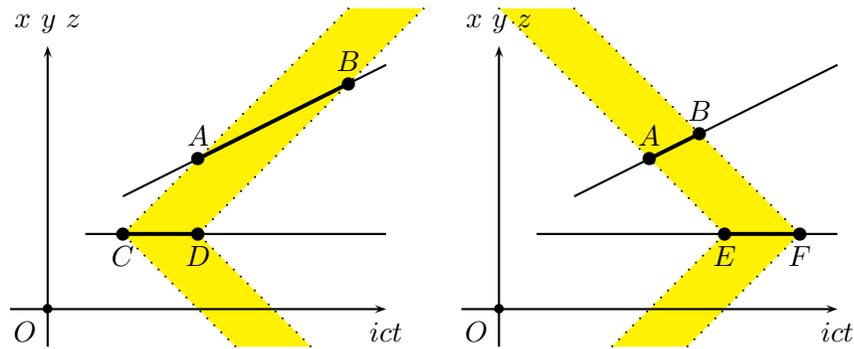
\begin{figure}[h!]
\begin{center}
\begin{pspicture}(-5.5,-1.5)(5.5,3)
\pspolygon*[linestyle=none,linecolor=yellow](-1,3)(-4,0)(-2.5,-1.5)(-1.5,-1.5)(-3,0)(0,3)
\psline{->}(-5.5,-1)(-0.5,-1)
\psline{->}(-5,-1.5)(-5,2.5)
\rput(-0.5,-1.3){$i c t$}
\rput(-5,2.8){$x \ y \ z$}
\psdot(-5,-1)
\rput(-5.3,-1.3){$O$}
\psline(-4.5,0)(-0.5,0)
\psline[showpoints=true,linewidth=1.5pt](-4,0)(-3,0)
\rput(-4,-0.3){$C$}
\rput(-3,-0.3){$D$}
\psline[linestyle=dotted](-4,0)(-1,3)
\psline[linestyle=dotted](-3,0)(0,3)
\psline[linestyle=dotted](-4,0)(-2.5,-1.5)
\psline[linestyle=dotted](-3,0)(-1.5,-1.5)
\psline(-4,0.5)(-0.5,2.25)
\psline[showpoints=true,linewidth=1.5pt](-3,1)(-1,2)
\rput(-3,1.3){$A$}
\rput(-1,2.3){$B$}
\pspolygon*[linestyle=none,linecolor=yellow](1,3)(4,0)(2.5,-1.5)(3.5,-1.5)(5,0)(2,3)
\psline{->}(0.5,-1)(5.5,-1)
\psline{->}(1,-1.5)(1,2.5)
\rput(5.5,-1.3){$i c t$}
\rput(1,2.8){$x \ y \ z$}
\psdot(1,-1)
\rput(0.7,-1.3){$O$}
\psline(1.5,0)(5.5,0)
\psline[showpoints=true,linewidth=1.5pt](4,0)(5,0)
\rput(4,-0.3){$E$}
\rput(5,-0.3){$F$}
\psline[linestyle=dotted](4,0)(1,3)
\psline[linestyle=dotted](5,0)(2,3)
\psline[linestyle=dotted](4,0)(2.5,-1.5)
\psline[linestyle=dotted](5,0)(3.5,-1.5)
\psline(2,0.5)(5.5,2.25)
\psline[showpoints=true,linewidth=1.5pt](3,1)(3.6666,1.3333)
\rput(3,1.3){$A$}
\rput(3.6666,1.6333){$B$}
\end{pspicture}
\caption{A test particle $AB$, in motion relative to the source particle, 
feels a retarded interaction from $CD$ and an advanced interaction from $EF$.}
\label{fig:fig4}
\end{center}
\end{figure}

The key observation is that, for two 
infinitesimal spacetime segments whose endpoints are connected by light signals,  
the ratio of their lengths depends on their orientation. 
For two particles at rest, the ratio is equal to one.
However, when one particle is at rest, but the other particle is in motion,
the ratio is no longer equal to one.
These two situations are represented graphically in Figures \ref{fig:fig3} and \ref{fig:fig4}.
We also notice that this ratio is a Lorentz invariant.
What really matters is the relative velocity.

The exact formula for the ratio of the lengths of the corresponding infinitesimal segments is derived in Appendix I. 
In general we have
\begin{equation}
\frac{AB}{CD} 
= \frac{(\mathbf{X}_2 - \mathbf{X}_{1 ret}) \cdot \mathbf{V}_{1 ret}}{(\mathbf{X}_2 - \mathbf{X}_{1 ret}) \cdot \mathbf{V}_2},
\label{eq:AB_CD_new}
\end{equation}
\begin{equation}
\frac{AB}{EF} 
= \frac{(\mathbf{X}_2 - \mathbf{X}_{1 adv}) \cdot \mathbf{V}_{1 adv}}{(\mathbf{X}_2 - \mathbf{X}_{1 adv}) \cdot \mathbf{V}_2},
\label{eq:AB_EF_new}
\end{equation}
where 
$\mathbf{X}_{1 ret}$ is the position four-vector of the retarded source particle $CD$,
$\mathbf{X}_{1 adv}$ is the position four-vector of the advanced source particle $EF$,
$\mathbf{X}_2$ is the position four-vector of the test particle $AB$,
$\mathbf{V}_{1 ret}$ is the four-velocity of the retarded source particle, 
$\mathbf{V}_{1 adv}$ is the four-velocity of the advanced source particle, and
$\mathbf{V}_2$ is the four-velocity of the test particle.

We work in the proper reference frame of the source particle at rest.
The four-velocity of the source particle (indexed with subscript 1) is
\begin{equation}
\mathbf{V}_{1 ret} = \mathbf{V}_{1 adv} = (\vec{0}, i c) = (0, 0, 0, i c).
\label{eq:rest_v1}
\end{equation}
The four velocity of the test particle (indexed with subscript 2), moving with radial velocity $v_x$ along the $x$ axis, is
\begin{equation}
\mathbf{V}_2 = (\gamma \vec{v}, i \gamma c) = (\gamma v_x, 0, 0, i \gamma c).
\label{eq:rest_v2}
\end{equation}
For the retarded interaction, the displacement four-vector is
\begin{equation}
\mathbf{X}_2 - \mathbf{X}_{1 ret} = (\vec{R}, i R) = (R, 0, 0, i R).
\label{eq:rest_ret}
\end{equation}
For the advanced interaction, the displacement four-vector is
\begin{equation}
\mathbf{X}_2 - \mathbf{X}_{1 adv} = (\vec{R}, - i R) = (R, 0, 0, - i R).
\label{eq:rest_adv}
\end{equation}
The ratios of the corresponding segments are calculated according to formulas (\ref{eq:AB_CD_new}) and (\ref{eq:AB_EF_new}).
For the retarded interaction
\begin{equation}
\frac{AB}{CD} 
= \frac{(R, 0, 0, i R) \cdot (0, 0, 0, i c)}{(R, 0, 0, i R) \cdot (\gamma v_x, 0, 0, i \gamma c)}
= \frac{- R c}{\gamma R v_x - \gamma R c}
= \frac{1}{\gamma \, (1 - v_x/c)},
\label{eq:galeriu1}
\end{equation}
and for the advanced interaction
\begin{equation}
\frac{AB}{EF} 
= \frac{(R, 0, 0, - i R) \cdot (0, 0, 0, i c)}{(R, 0, 0, - i R) \cdot (\gamma v_x, 0, 0, i \gamma c)}
= \frac{R c}{\gamma R v_x + \gamma R c}
= \frac{1}{\gamma \, (1 + v_x/c)}.
\label{eq:galeriu2}
\end{equation}

To make further progress with our theory 
of time symmetric action-at-a-distance electrodynamics, we recall an observation made by 
Olivier Costa de Beauregard \cite{costa}. 
The equations of motion of a relativistic point particle and the equations describing
a classical elastic string in static equilibrium are isomorphic (of similar form and structure). 
We can think of the worldline of a particle as an elastic string at rest in spacetime.
This isomorphism was also independently rediscovered in Ref. \cite{galeriuAAAD}.

A relativistic point particle with four-momentum $\mathbf{P}$, subject to a four-force $\mathbf{F}$, undergoes
a trajectory described by 
\begin{equation}
\mathbf{P}_B - \mathbf{P}_A = \mathbf{F} \, d\tau,
\label{eq:worldline}
\end{equation}
where $A$ and $B$ are two infinitesimally close points on the worldline of the particle, 
separated by a proper time interval $d\tau$.
The four-momentum $\mathbf{P}$ is a time-like vector directed towards the future.

A static classical elastic string under a tension of magnitude $T$, and subject to a linear force density $\mathbf{f}$, 
has segments of infinitesimal length $ds$ in static equilibrium.
We give the tension vector $\mathbf{T}$
the direction that points towards the future.
The vector $\mathbf{T}$ is tangent to the string. 
The static equilibrium condition for the string segment $AB$ of length $ds$ becomes 
\begin{equation}
\mathbf{T}_B + (- \mathbf{T}_A) + \mathbf{f} \, ds = 0, 
\label{eq:string}
\end{equation}
where the minus sign indicates that the tension forces
acting at the end points of the infinitesimal segment $AB$ have opposite directions. 

\begin{figure}[h!]
\begin{center}
\begin{pspicture}(0,0)(13,6)
\psline{->}(0,1.5)(6,1.5)
\psline{->}(0.5,1)(0.5,5.5)
\psdot(0.5,1.5)
\rput(0.2,1.2){$O$}
\rput(0.5,5.8){$x \ y \ z$}
\rput(6,1.2){$ict$}
\pscurve(0.5,2.5)(2,3.25)(3,3.4)(4,3.25)(5.5,2.5)
\psdot(2,3.25)
\rput(2,2.9){$A$}
\psdot(4,3.25)
\rput(4,2.9){$B$}
\psline[linewidth=1.6pt]{->}(2,3.25)(4,4)
\rput(4,4.3){$\mathbf{P}_A$}
\psline[linewidth=1.6pt]{->}(4,3.25)(6,2.5)
\rput(6,2.9){$\mathbf{P}_B$}
\psline[linewidth=1.6pt]{->}(3,3.4)(3,1.9)
\rput(3.3,1.9){$\mathbf{F}$}
\rput(3,0.2){$(a)$}
\psline{->}(7,1.5)(13,1.5)
\psline{->}(7.5,1)(7.5,5.5)
\psdot(7.5,1.5)
\rput(7.2,1.2){$O$}
\rput(7.5,5.8){$x \ y \ z$}
\rput(13,1.2){$ict$}
\pscurve(7.5,2.5)(9,3.25)(10,3.4)(11,3.25)(12.5,2.5)
\psdot(9,3.25)
\rput(9,2.9){$A$}
\psdot(11,3.25)
\rput(11,2.9){$B$}
\psline[linewidth=1.6pt]{->}(9,3.25)(7,2.5)
\rput(7,2.9){$\mathbf{- T_A}$}
\psline[linewidth=1.6pt]{->}(11,3.25)(13,2.5)
\rput(13,2.9){$\mathbf{T_B}$}
\psline[linewidth=1.6pt]{->}(10,3.4)(10,4.9)
\rput(10.3,4.9){$\mathbf{f}$}
\rput(10,0.2){$(b)$}
\end{pspicture}
\caption{The worldline of a relativistic point particle (a)
can be looked upon as a spacetime string in static equilibrium (b).}
\label{fig:fig2}
\end{center}
\end{figure}
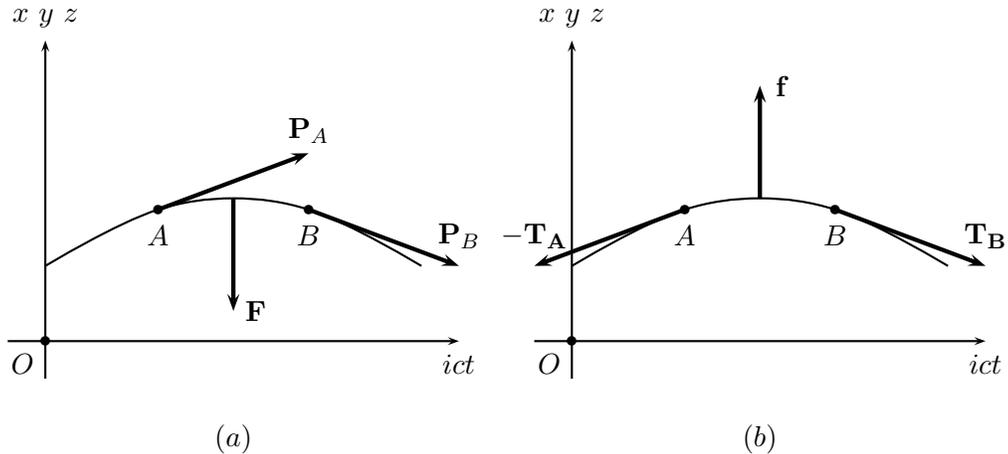

Equations (\ref{eq:worldline}) and (\ref{eq:string}) are equivalent, provided that the length of the 
worldline segment $AB$ is
$ds = i \, c\, d\tau$, 
the tension in the string is $\mathbf{T} = \mathbf{P} / \tau_o$, 
and the linear force density is $\mathbf{f} = - \mathbf{F} / s_o$. 
Here $\tau_o$ is a yet undetermined real constant quantity with units of time, introduced for dimensional reasons,
and $s_o = i \, c\, \tau_o$. 
We notice that the four-force acting on the material point particle and the linear four-force density acting on the
infinitesimal worldline segment have opposite directions, as seen from Fig. \ref{fig:fig2}. 

According to the spacetime string model described so far, when both electrically charged 
particles are at rest, as considered in Subsection 2.1, 
the retarded and the advanced linear four-force densities are
\begin{eqnarray}
\mathbf{f}_{ret} = \frac{-1}{s_o} \mathbf{F}_{ret}
= \frac{-1}{s_o} \Big(\frac{Q q}{2 R^2}, 0, 0, i \frac{Q q}{2 R^2}\Big), \label{eq:0_0_ret} \\
\mathbf{f}_{adv} = \frac{-1}{s_o} \mathbf{F}_{adv}
= \frac{-1}{s_o} \Big(\frac{Q q}{2 R^2}, 0, 0, - i \frac{Q q}{2 R^2}\Big). \label{eq:0_0_adv}
\end{eqnarray} 

We are now ready to describe, in a geometrical manner, the velocity dependence 
of the four-force (\ref{eq:fourforce}). 
By postulate (Postulate II) we assume that, 
in the inertial reference frame where the source particle is instantaneously at rest,
the linear four-force density acting on the test particle is proportional to the product of the
Coulombian electrostatic force with the ratio of 
the lenght of the infinitesimal segment on the worldline of the test particle to 
the length of the corresponding infinitesimal segment on the worldline of the source particle \cite{arcane}. 

For a test particle at rest, $AB/CD = 1$ and $AB/EF = 1$, which means that equations (\ref{eq:0_0_ret})-(\ref{eq:0_0_adv}), 
and therefore also equations (\ref{equation1})-(\ref{equation3}), stay the same.

However, for a test particle with only radial velocity, the retarded and the advanced linear four-force densities become
\begin{multline}
\mathbf{f}_{ret} = \frac{-1}{s_o} \Big(\frac{Q q}{2 R^2}, 0, 
0, i \frac{Q q}{2 R^2}\Big) \frac{AB}{CD} \\
= \frac{-1}{s_o} \Big(\frac{Q q}{2 R^2} \frac{1}{\gamma \, (1 - v_x/c)}, 0, 
0, i \frac{Q q}{2 R^2} \frac{1}{\gamma \, (1 - v_x/c)}\Big), 
\label{eq:0_0_ret2}
\end{multline}
\begin{multline}
\mathbf{f}_{adv} = \frac{-1}{s_o} \Big(\frac{Q q}{2 R^2}, 0, 
0, - i \frac{Q q}{2 R^2}\Big) \frac{AB}{EF} \\
= \frac{-1}{s_o} \Big(\frac{Q q}{2 R^2} \frac{1}{\gamma \, (1 + v_x/c)}, 0, 
0, - i \frac{Q q}{2 R^2} \frac{1}{\gamma \, (1 + v_x/c)}\Big).
\label{eq:0_0_adv2}
\end{multline} 
Due to the \lq\lq magic\rq\rq\ fact that
\begin{eqnarray}
\frac{1}{(1 - v_x/c)} + \frac{1}{(1 + v_x/c)} = \frac{2}{1 - v_x^2/c^2} = 2 \gamma^2, \label{eq17} \\
\frac{1}{(1 - v_x/c)} - \frac{1}{(1 + v_x/c)} = \frac{2 v_x/c}{1 - v_x^2/c^2} = 2 \gamma^2 \frac{v_x}{c}, \label{eq18}
\end{eqnarray}
the total linear four-force density becomes
\begin{equation}
\mathbf{f} = \mathbf{f}_{ret} + \mathbf{f}_{adv} 
= \frac{-1}{s_o} \Big(\frac{Q q}{R^2} \gamma, 0, 0, i \frac{Q q}{R^2} \gamma \frac{v_x}{c}\Big),
\label{eq19}
\end{equation}
and the total four-force $\mathbf{F} = - s_0 \, \mathbf{f}$
is in full agreement with equation (\ref{eq:fourforce}).

Please keep in mind that equations (\ref{eq17})-(\ref{eq18}) hold only when 
the velocity of the test particle is purely radial, that means
parallel to the electric field of the stationary source particle.

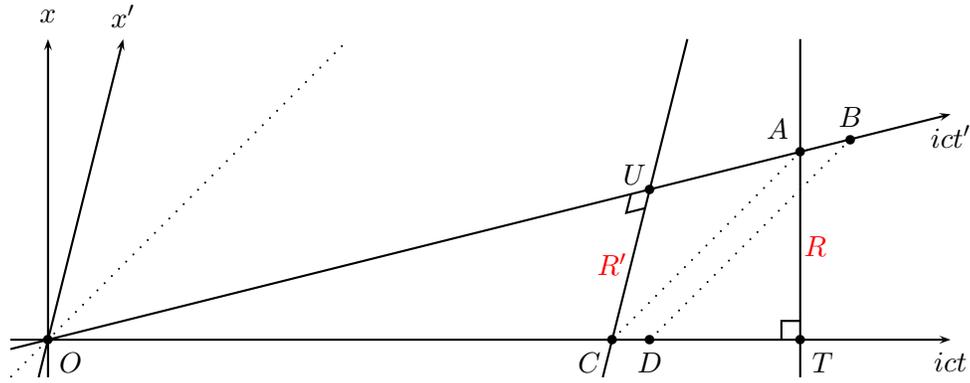
\begin{figure}[h!]
\begin{center}
\begin{pspicture}(-0.5,-0.5)(12,4.5)
\psline{->}(-0.5,0)(12,0)
\psline{->}(0,-0.5)(0,4)
\psdot(0,0)
\rput(0.3,-0.3){$O$}
\rput(12,-0.3){$i c t$}
\rput(0,4.3){$x$}
\psline{->}(-0.5,-0.125)(12,3)
\psline{->}(-0.125,-0.5)(1,4)
\rput(12,2.7){$i c t'$}
\rput(1,4.3){$x'$}
\psline[linestyle=dotted](-0.5,-0.5)(4,4)
\psline(7.375,-0.5)(8.5,4)
\psdot(7.5,0)
\rput(7.2,-0.3){$C$}
\psdot(8,2)
\rput(7.8,2.2){$U$}
\rput(7.5,1){\textcolor{red}{$R'$}}
\psline(10,-0.5)(10,4)
\psdot(10,0)
\rput(10.3,-0.3){$T$}
\psdot(10,2.5)
\rput(9.7,2.8){$A$}
\rput(10.2,1.25){\textcolor{red}{$R$}}
\psline[linestyle=dotted](7.5,0)(10,2.5)        
\psline[linestyle=dotted](8,0)(10.6667,2.6667)      
\psdot(8,0)
\rput(8,-0.3){$D$}
\psdot(10.6667,2.6667)
\rput(10.6667,2.9667){$B$}
\psline(9.75,0)(9.75,0.25)(10,0.25)
\psline(7.75,1.9375)(7.6875,1.6875)(7.9375,1.75)
\end{pspicture}
\caption{When the relative velocity is along the radial direction,
the interacting particles $AB$ and $CD$ are in the same Minkowski plane.}
\label{fig:fig5}
\end{center}
\end{figure}

A geometrical derivation of formulas (\ref{eq:galeriu1}) and (\ref{eq:galeriu2}),
for the case of only radial motion, was given in Ref. \cite{galeriuAAAD}.
Here we give a different geometrical derivation, as a preliminary step towards a discussion of the relationship between 
the four-forces of action and reaction.
Since under a time reversal operation the velocity $\vec{v}$ of the test particle changes sign, 
and the formulas (\ref{eq:galeriu1}) and (\ref{eq:galeriu2}) turn into each other, it is enough to  demonstrate just one of them.
In particular, we assume that $v_x > 0$ and we calculate the ratio $AB/CD$ associated with the retarded interaction.
Due to the purely radial motion we can also assume that 
the origin of the proper reference frame $K$ of the source particle 
and the origin of the proper reference frame $K'$ of the test particle  
coincide when $t = t'= 0$ .
Each electrically charged particle is at rest at the origin of its proper reference frame.
Their worldlines are the $i c t$ and $i c t'$ time axes.
This situation is represented graphically in Fig. \ref{fig:fig5}.

From point $A$ we draw a line perpendicular to the $i c t$ axis, which intersects it at point $T$. 
The length of segment $AT$ is $R$.
From point $C$ we draw a line perpendicular to the $i c t'$ axis, which intersects it at point $U$. 
The length of segment $CU$ is $R'$.
Segment $AC$ has a null length. 
As a consequence, the length of segment $CT$ is $i R$ and the length of segment $UA$ is $i R'$.
Since segments $CA$ and $DB$ are parallel, due to the theorem of Thales we have
\begin{equation}
\frac{AB}{CD} = \frac{OA}{OC}.
\label{eq:Thales}
\end{equation}

Let $t$ be the time coordinate of point $A$ in the proper reference frame of the source charge. It follows that
\begin{eqnarray}
AT = v_x \, t, \\
OT = i c t, \\
OA = \sqrt{OT^2 + AT^2} = i \sqrt{c^2 - v_x^2} \, t, \\
CT = i \, AT = i v_x t, \\
OC = OT - CT = i c t - i v_x \, t, \\
\frac{AB}{CD} = \frac{OA}{OC} 
= \frac{ i \sqrt{c^2 - v_x^2} \, t}{i c t - i v_x \, t} = \frac{\sqrt{c^2 - v_x^2}}{c - v_x}  
= \frac{1}{\gamma \, (1 - v_x/c)},
\end{eqnarray}
and this concludes the geometrical proof of formula (\ref{eq:galeriu1}).

We want to verify that the retarded four-force acting on segment $AB$, due to segment $CD$,
has the same magnitude and opposite direction as the advanced four-force acting on segment $CD$, due to segment $AB$.
We find each four-force by multiplying the relevant linear four-force density by the length of the 
worldline segment on which it acts.
According to the mechanism described, 
in the proper reference frame $K$ of particle $CD$ the retarded four-force acting on particle $AB$ is
\begin{equation}
\mathbf{f}_{ret} \, AB = \frac{-1}{s_o} \Big( \frac{Q q}{2 R^2} \frac{AB}{CD} AB, 0, 0, i \frac{Q q}{2 R^2} \frac{AB}{CD} AB \Big).
\label{fretAB}
\end{equation}
Similarly, in the proper reference frame $K'$ of particle $AB$ the advanced four-force acting on particle $CD$ is
\begin{equation}
\mathbf{g}_{adv} \, CD = \frac{-1}{s_o} \Big( - \frac{Q q}{2 R'^2} \frac{CD}{AB} CD, 0, 0, - i \frac{Q q}{2 R'^2} \frac{CD}{AB} CD \Big).
\end{equation}
The easiest way to convert the four-force components from $K'$ to $K$ is to notice that in reference frame $K'$
\begin{equation}
\overrightarrow{CA} = \overrightarrow{CU} + \overrightarrow{UA} 
= (R', 0, 0, 0) + (0, 0, 0, i R')
= (R', 0, 0, i R'),
\end{equation}
while in reference frame $K$
\begin{equation}
\overrightarrow{CA} = \overrightarrow{CT} + \overrightarrow{TA} 
= (0, 0, 0, i R) + (R, 0, 0, 0)
= (R, 0, 0, i R).
\end{equation}
The components in $K$ of such a vector of null length are equal to the components in $K'$ multiplied by $R/R'$.
Accordingly, in the proper reference frame $K$ of particle $CD$ the advanced four-force acting on particle $CD$ is
\begin{equation}
\mathbf{g}_{adv} \, CD = \frac{-1}{s_o} \Big( - \frac{Q q}{2 R'^2} \frac{CD}{AB} CD \frac{R}{R'}, 0, 
0, - i \frac{Q q}{2 R'^2} \frac{CD}{AB} CD \frac{R}{R'} \Big).
\label{favdCD}
\end{equation}

Comparing the four-force expressions (\ref{fretAB}) and (\ref{favdCD}), 
we see that the principle of action and reaction is verified if and only if 
\begin{equation}
\frac{1}{R^2} \frac{AB}{CD} AB = \frac{1}{R'^2} \frac{CD}{AB} CD \frac{R}{R'},
\label{eq:PARcondition}
\end{equation}
an equation that simplifies to
\begin{equation}
\frac{AB}{CD} = \frac{R}{R'}.
\label{eq:PARcondition2}
\end{equation}
Due to equation (\ref{eq:Thales}), condition (\ref{eq:PARcondition2}) is equivalent to
\begin{equation}
\frac{OA}{OC} = \frac{AT}{CU},
\end{equation}
an equation that follows from the similarity of  $\triangle OTA$ and $\triangle OUC$.
An algebraic derivation of equation (\ref{eq:PARcondition2}) is given in Appendix B.
We also notice that equation (\ref{eq:PARcondition2}) is related to the surface area of $ABDC$, which,
up to the first order in the infinitesimals, is equal to $AB \times R' = CD \times R$.

As a side note, as seen from equation (\ref{eq:PARcondition}), 
Postulate II together with the principle of action and reaction 
can explain why the electrostatic force has an inverse square dependence on the distance
between charges.

\subsection{Test particle with non-radial velocity $\vec{v} = (v_x, v_y, 0)$}

When the test particle has a non-radial velocity $\vec{v} = (v_x, v_y, 0)$,
the speed is $v = \sqrt{v_x^2 + v_y^2}$ 
and the Lorentz factor is $\gamma = 1 / \sqrt{1 - (v_x^2 + v_y^2) / c^2}$.

With this new value of the Lorentz factor,
the four-force has the same expression (\ref{eq:fourforce}),
and the ratios of corresponding segments have the same expressions (\ref{eq:galeriu1}) and (\ref{eq:galeriu2}).
A geometrical derivation of formulas (\ref{eq:galeriu1}) and (\ref{eq:galeriu2}),
for the case of non-radial motion, was also given in Ref. \cite{galeriuAAAD}.

In addition to the algebraic derivation from Appendix B, 
we give here an alternative geometrical proof of the fact that, for the retarded interaction, 
the value of $R / R'$ is equal to the value of $AB / CD$ from equation (\ref{eq:galeriu1}).

We select a reference frame in which the retarded source particle at $C$ is at rest at the origin.
At time $t$ the test particle is at point $A$ on the $Ox$ axis. The $Ox$ axis represents the radial direction.
The origin on the time axis is chosen in such a way that,
at the initial time $t_0 = 0$, the test particle is at point $W$ on the $Oy$ axis. 
This situation is represented graphically in Fig. \ref{fig:fig6}.

\begin{figure}[h!]
\begin{center}
\begin{pspicture}(-1,-1)(12,4.5)
\psline{->}(-0.5,0)(12,0)
\psline{->}(0,-0.5)(0,4)
\psdot(0,0)
\rput(0.3,-0.3){$O$}
\rput(12,-0.3){$i c t$}
\rput(0,4.3){$x$}
\psline{->}(4,3)(-1,-0.75)
\rput(-1,-1){$y$}
\psline{->}(0.4,1.3)(11.6,2.7)
\rput(11.6,2.4){$i c t'$}
\psdot(2,1.5)
\rput(1.8,1.8){$W$}
\psline(7.3889,-0.5)(8.25,3.375)          
\psdot(7.5,0)
\rput(7.2,-0.3){$C$}
\psdot(8,2.25)
\rput(7.8,2.45){$U$}
\rput(7.5,1.1){\textcolor{red}{$R'$}}
\psline(10,-0.5)(10,4)
\psdot(10,0)
\rput(10.3,-0.3){$T$}
\psdot(10,2.5)
\rput(9.7,2.8){$A$}
\rput(10.2,1.25){\textcolor{red}{$R$}}
\psline[linestyle=dotted](7.5,0)(10,2.5)        
\psline(9.75,0)(9.75,0.25)(10,0.25)       
\psline(8.05423,2.49405)(8.3023,2.52506)(8.24807,2.28101)       
\psline(2,1.5)(7.5,0)         
\psline(0,0)(10,2.5)         
\end{pspicture}
\caption{When the relative velocity is not along the radial direction,
the worldlines of the two interacting particles, 
the $i c t$ and $i c t'$ time axes, are no longer in the same Minkowski plane.}
\label{fig:fig6}
\end{center}
\end{figure}

It follows that
\begin{eqnarray}
AT = v_x \, t = R, \\
OW = v_y \, t, \\
OT = i c t, \\
OA = \sqrt{AT^2 + OT^2} = \sqrt{v_x^2 - c^2} \, t = i \sqrt{c^2 - v_x^2} \, t,\\
CT = i \, AT = i v_x \, t, \\
OC = OT - CT = i c t - i v_x \, t = i (c - v_x) t, \\
WC = \sqrt{OW^2 + OC^2} = \sqrt{v_y^2 - (c - v_x)^2} \, t = i \sqrt{(c - v_x)^2 - v_y^2} \, t, \\
WA = \sqrt{OW^2 + OA^2} = \sqrt{v_y^2 + v_x^2 - c^2} \, t = i \sqrt{c^2 - v_x^2 - v_y^2} \, t, \\
CU = R', \\
UA = i \, CU = i R', \\
WU = WA - UA = i \sqrt{c^2 - v_x^2 - v_y^2} \, t - i R'.
\end{eqnarray}
In order to find an expression for $R'$ we write the Pythagorean theorem in $\triangle WUC$
\begin{equation}
WC^2 = WU^2 + CU^2.
\end{equation}
We solve the resulting equation
\begin{equation}
\Big(i \sqrt{(c - v_x)^2 - v_y^2}\, t\Big)^2 = \Big(i \sqrt{c^2 - v_x^2 - v_y^2} \, t - i R'\Big)^2 + (R')^2, \\
\end{equation}
and we find, as expected, that
\begin{equation}
R' = \frac{v_x \, (c - v_x) t}{\sqrt{c^2 - v_x^2 - v_y^2}} = \frac{c - v_x}{\sqrt{c^2 - v_x^2 - v_y^2}} R = \gamma \Big(1 - \frac{v_x}{c}\Big) R.
\end{equation}

As an added benefit, this geometrical proof of the formula for the $R/R'$ ratio also allows us to discover 
the surprising fact that $\triangle WAC$ and $\triangle OAC$ have the same area.
Indeed
\begin{equation}
WA \times CU = i \sqrt{c^2 - v_x^2 - v_y^2} \, t \times \frac{v_x \, (c - v_x) t}{\sqrt{c^2 - v_x^2 - v_y^2}} = OC \times AT.
\end{equation}
How can this be, since $\triangle OAC$ is the projection onto plane $x O i c t$ of $\triangle WAC$?
The well known results from Euclidean geometry do not apply here, in Minkowski space,
because in this specific example the two Minkowski planes intersect along a null line.

Due to the new value of the Lorentz factor, the last equalities in equations (\ref{eq17}) and (\ref{eq18}) no longer hold.
When adding equations (\ref{eq:0_0_ret2}) and (\ref{eq:0_0_adv2}), instead of equation (\ref{eq19}) we now obtain
\begin{equation}
\mathbf{f} = \mathbf{f}_{ret} + \mathbf{f}_{adv} 
= \frac{-1}{s_o} \Big(\frac{Q q}{R^2} \gamma, 0, 0, i \frac{Q q}{R^2} \gamma \frac{v_x}{c}\Big) \frac{1}{\gamma^2 \, (1 - v_x^2/c^2)}.
\label{eq35}
\end{equation}

We notice that there is an extra factor
\begin{equation}
\Gamma = \frac{1}{\gamma^2 \, (1 - v_x^2/c^2)} 
= \frac{c^2 - v^2}{c^2 - v_x^2}
= \frac{c^2 - v_x^2 - v_y^2}{c^2 - v_x^2}
\label{eq:Gammafactor}
\end{equation}
that makes our expression of the four-force a little bit different from what we expect from classical electrodynamics.
Formula (\ref{eq:Gammafactor}) shows that only $v_y$, the non-radial component of the velocity of the test particle, 
perpendicular to the radial electric field $\vec{E} = (E_x, 0, 0)$ of the source particle at rest, 
can make the $\Gamma$ factor deviate from a unit value.

From an experimental point of view, the velocity of electrons in metals is not relativistic, 
and the correction due to the $\Gamma$ factor is very small.
In linear particle accelerators the velocity of the electric charges is parallel to the electric field, 
and this is just like the case of motion in the radial direction.  
In synchrotons relativistic electrons or protons are held in circular orbits not by electrostatic fields,
but by magnetic fields.

From a theoretical point of view, as was noticed by Gen Yoneda in Appendix A of Ref. \cite{GenYoneda},
we expect to see some disagreements between classical electrodynamics and an alternative theory that 
implements \lq\lq the parallel condition\rq\rq.

\subsection{Test particle with non-radial velocity $\vec{v} = (v_x, v_y, v_z)$}

Consider a source particle with electric charge $Q$ at rest at the origin $(0, 0, 0)$ of the reference frame,
and a test particle with electric charge $q$ at the position given by $\vec{R} = (R_x, R_y, R_z)$.
The distance between the two particles is $R = \sqrt{R_x^2 + R_y^2 + R_z^2}$.
The test particle has velocity $\vec{v} = (v_x, v_y, v_z)$ and
the Lorentz factor is $\gamma = 1 / \sqrt{1 - v^2 / c^2} = 1 / \sqrt{1 - (v_x^2 + v_y^2 + v_z^2) / c^2}$.

The ratios of the corresponding segments are calculated according to formulas (\ref{eq:AB_CD_new}) and (\ref{eq:AB_EF_new}).
For the retarded interaction
\begin{equation}
\frac{AB}{CD} 
= \frac{(\vec{R}, i R) \cdot (\vec{0}, i c)}{(\vec{R}, i R) \cdot (\gamma \vec{v}, i \gamma c)}
= \frac{- R c}{\gamma \vec{R} \cdot \vec{v} - \gamma R c}
= \frac{1}{\gamma \, (1 - v_{rad}/c)},
\label{eq:galeriu1xyz}
\end{equation}
and for the advanced interaction
\begin{equation}
\frac{AB}{EF} 
= \frac{(\vec{R}, - i R) \cdot (\vec{0}, i c)}{(\vec{R}, - i R) \cdot (\gamma \vec{v}, i \gamma c)}
= \frac{R c}{\gamma \vec{R} \cdot \vec{v} + \gamma R c}
= \frac{1}{\gamma \, (1 + v_{rad}/c)},
\label{eq:galeriu2xyz}
\end{equation}
where by definition the radial component of the velocity is $v_{rad} = \vec{v} \cdot \vec{R} / R$.
We also have a non-radial component, such that $v^2 = v_{rad}^2 + v_{nonrad}^2$.

The retarded and the advanced linear four-force densities become
\begin{equation}
\mathbf{f}_{ret} = \frac{-1}{s_o} \Big(\frac{Q q}{2 R^2} \frac{\vec{R}}{R} \frac{1}{\gamma \, (1 - v_{rad}/c)}, 
i \frac{Q q}{2 R^2} \frac{1}{\gamma \, (1 - v_{rad}/c)}\Big), 
\label{eq:0_0_ret3}
\end{equation}
\begin{equation}
\mathbf{f}_{adv} = \frac{-1}{s_o} \Big(\frac{Q q}{2 R^2} \frac{\vec{R}}{R} \frac{1}{\gamma \, (1 + v_{rad}/c)}, 
- i \frac{Q q}{2 R^2} \frac{1}{\gamma \, (1 + v_{rad}/c)}\Big),
\label{eq:0_0_adv3}
\end{equation} 
and the total linear four-force density becomes
\begin{equation}
\mathbf{f} = \mathbf{f}_{ret} + \mathbf{f}_{adv} 
= \frac{-1}{s_o} \Big(\frac{Q q}{R^2} \frac{\vec{R}}{R} \gamma, i \frac{Q q}{R^2} \gamma \frac{v_{rad}}{c}\Big) 
\frac{1}{\gamma^2 \, (1 - v_{rad}^2/c^2)}.
\label{eq:55xyz}
\end{equation}

We notice that there is an extra factor
\begin{equation}
\Gamma = \frac{1}{\gamma^2 \, (1 - v_{rad}^2/c^2)} 
= \frac{c^2 - v^2}{c^2 - v_{rad}^2}
= \frac{c^2 - v_{rad}^2 - v_{nonrad}^2}{c^2 - v_{rad}^2}
\end{equation}
that makes our expression of the four-force
\begin{equation}
\mathbf{F} = - s_0 \, \mathbf{f} 
= \Big(\frac{Q q}{R^2} \frac{\vec{R}}{R} \gamma \Gamma, i \frac{Q q}{R^2} \gamma \frac{v_{rad}}{c} \Gamma\Big)
\label{eq:60fourforce}
\end{equation}
a little bit different from what we expect from classical electrodynamics.
The four-force (\ref{eq:60fourforce})
is what we get when the Coulombian force $\vec{F} = Q q \vec{R} / R^3$ in equation (\ref{eq:fourforce}) is multiplied by $\Gamma$.
The four-force (\ref{eq:60fourforce}) is still orthogonal to the four-velocity of the test particle.

Although the extra $\Gamma$ factor is unexpected, we cannot simply eliminate it by postulate, 
since that would destroy the balance between
action and reaction. As proven in Appendix C, the two $\Gamma$ factors for action and reaction are equal to each other 
only when each particle, relative to the other particle, is moving only in the radial direction.

\section{Source particle in hyperbolic motion, and field point that is
simultaneous with the center of the hyperbola}

Consider a source particle with electric charge $Q$, moving in hyperbolic motion along the $x$ axis.
By our choice, the origin $O$ of the 4D reference frame is also the center of the hyperbola.
The hyperbola intersects the positive $x$ axis in a point at a distance $a$ from the origin.
Let $s$ be the arclength on the hyperbola, related to the proper time $\tau$ of the particle by
the formula $ds = i \, c \, d\tau$. In analogy with the definition of the value in radians of an angle
in 3D Euclidean space, we introduce the imaginary angle $\psi$ based on 
the formula $ds = a \, d\psi$ \cite{GaleriuMS}. It follows that $d\psi / d\tau = i \, c / a$.
The positive direction of the angle coordinate $\psi$ points into the future.
The worldline of the source particle is described by the position four-vector
\begin{equation}
\mathbf{X}_1 = \Big(a \, \cos(\psi), 0, 0, a \, \sin(\psi)\Big),
\label{eq:4position}
\end{equation}
and the four-velocity of the source particle is given by
\begin{equation}
\mathbf{V}_1 = \frac{d\mathbf{X}_1}{d\tau}
= \Big(- i \, c \, \sin(\psi), 0, 0, i \, c \, \cos(\psi)\Big).
\label{eq:4velocity}
\end{equation}

Consider a field point $A$ that is simultaneous with the center $O$ of the hyperbola,
in a given inertial reference frame $K$.
The position four-vector of point $A$ is
\begin{equation}
\mathbf{X}_2 = \Big(\rho, y, z, 0\Big),
\label{eq:4positionP}
\end{equation}

Relative to field point $A$, there is a retarded source charge at point $C$ and 
an advanced source charge at point $E$. 
The displacement four-vector from the source particle to the field point is
\begin{equation}
\mathbf{X}_2 - \mathbf{X}_1 = \Big(\rho - a \, \cos(\psi), y, z, - a \, \sin(\psi)\Big).
\end{equation}
For both the retarded and the advanced electromagnetic interactions
the condition $(\mathbf{X}_2 - \mathbf{X}_1) \cdot (\mathbf{X}_2 - \mathbf{X}_1) = 0$
reduces to the equation
\begin{equation}
\cos(\psi) = \frac{\rho^2 + y^2 + z^2 + a^2}{2 a \rho},
\label{eq:costheta}
\end{equation}
which has two imaginary solutions, $\theta$ and $- \theta$, that have the same absolute value.
For the retarded solution $\psi_{ret} = - \theta$ is a negative imaginary angle,
while for the advanced solution $\psi_{adv} = \theta$ is a positive imaginary angle.

\begin{figure}[h!]
\begin{center}
\begin{pspicture}(-4,-1)(4,6.5)
\psaxes[ticks=none,labels=none]{->}(0,0)(-4,-1)(4,6)[$i c t$,0][$x$,90]
\rput(-0.25,-0.25){$O$}
\psline[linestyle=dotted](-1,-1)(4,4)
\psline[linestyle=dotted](1,-1)(-4,4)
\psparametricplot[plotstyle=curve,linewidth=1.5pt]{-1.1}{1.1}{t SINH 3 mul t COSH 3 mul}
\psline[showpoints=true](0,0)(-1.232,3.243)       
\rput(-1.462,3.023){$C$}
\psarc(0,0){0.6}{90}{110.8}
\rput(-0.14,0.8){$\theta$}
\psarc(0,0){0.7}{69.2}{90}
\rput(0.17,0.9){$\theta$}
\psdot(0,5)
\rput(0.25,5.25){$S$}
\pspolygon*(0,5)(-1.04,5.46)(-0.96,5.54)          
\rput(-1.2,5.7){$A$}
\psline(-1.232,3.243)(0,3.243)
\psline[showpoints=true](0,3.243)(1.232,3.243)      
\rput(0.24,3.463){$M$}
\psline[showpoints=true](0,0)(1.232,3.243)       
\rput(1.452,3.023){$E$}
\psline[linestyle=dotted](-1.232,3.243)(-1,5.5)       
\psline[linestyle=dotted](1.232,3.243)(-1,5.5)        
\pspolygon*(0,3.243)(-1.04,5.46)(-0.96,5.54)          
\psdot(-1,5.5)
\end{pspicture}
\caption{In the reference frame $K$ where the field point $A$ and the center of the hyperbola
are simultaneous, the electric field has the direction of $\overrightarrow{MA}$.}
\label{fig:fig7}
\end{center}
\end{figure}
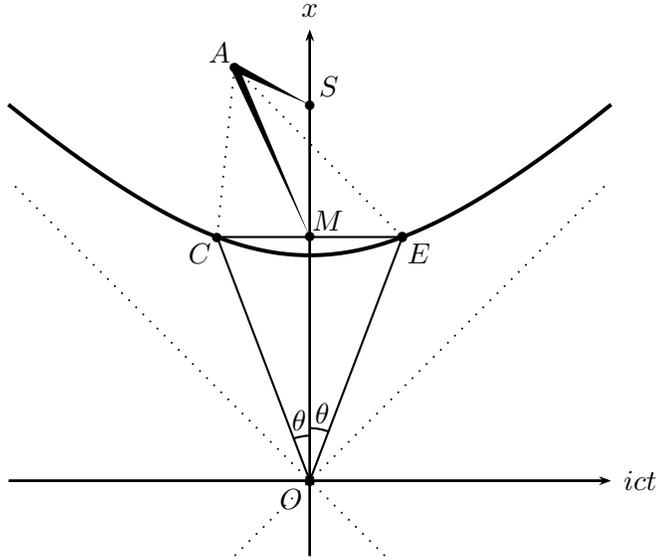

Let $M$ be the intersection of line $CE$ with the $x$ axis, as seen in Fig. \ref{fig:fig7}.
Since $OC = OE = a$, $\triangle OCE$ is an isosceles triangle and the angle bisector $OM$
is also a median and a height. The {\bf direction of the electric field} at point $A$,
as shown in Ref. \cite{GaleriuHYP}, is the direction of the displacement four-vector 
$\overrightarrow{MA}$.
This direction reveals the time symmetric nature of the interaction, because \cite{arcane}
\begin{equation}
\overrightarrow{MA} = \frac{1}{2} \, \overrightarrow{CA} + \frac{1}{2} \, \overrightarrow{EA}.
\end{equation}
Let $S$ be the projection of point $A$ on the $x$ axis.
Since the time axis is perpendicular to $AO$ (by hypothesis), 
and since the time axis is also perpendicular to the $x$ axis,
it follows that $AS$ is perpendicular to the time axis,
which means that $AS$ is perpendicular to the $x O i c t$ Minkowski plane.

We have \cite{GaleriuHYP}
\begin{eqnarray}
OC = OE = a, \\
OM = OC \cos(\theta) = a \cos(\theta), \\
CM = ME = OC \sin(\theta) = a \sin(\theta), \\
OS = \rho, \\
AS = \sqrt{y^2 + z^2}, \\
MS = OS - OM = \rho - a \cos(\theta), \\
MA = \sqrt{MS^2 + AS^2} = \sqrt{\rho^2 + a^2 \cos^2(\theta) - 2 \rho a \cos(\theta) + y^2 + z^2}.
\end{eqnarray}
With the help of equation(\ref{eq:costheta}) we find that
\begin{equation}
MA = \sqrt{a^2 \cos^2(\theta) - a^2} = a \sqrt{\cos^2(\theta) - 1} = - i a \sin(\theta). 
\end{equation}

We introduce the notation $\overrightarrow{MA} = (\vec{R}, 0)$, 
meaning that in reference frame $K$ we have
\begin{equation}
\vec{R} = \Big(\rho - a \cos(\theta), y, z\Big).
\end{equation}
This 3D vector with the direction of the electric field
is the equivalent of the radial position vector when the source charge was at rest.
The distance
\begin{equation}
R = - i a \sin(\theta)
\end{equation}
is a Lorentz invariant quantity.

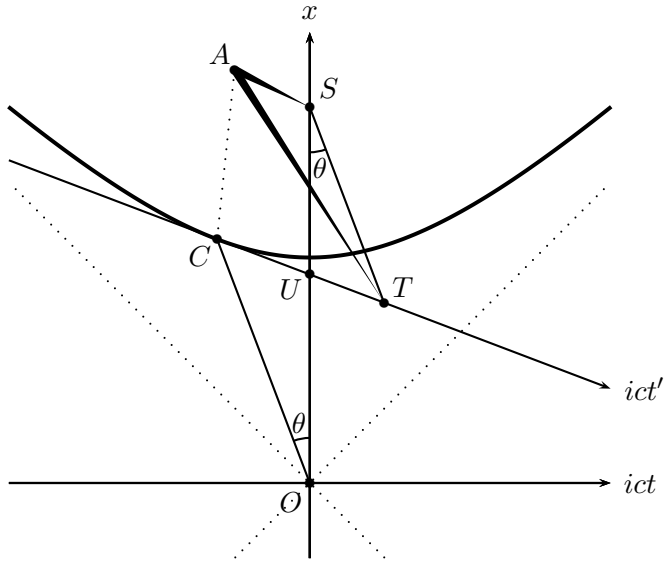
\begin{figure}[h!]
\begin{center}
\begin{pspicture}(-4,-1)(4,6.5)
\psaxes[ticks=none,labels=none]{->}(0,0)(-4,-1)(4,6)[$i c t$,0][$x$,90]
\rput(-0.25,-0.25){$O$}
\psline[linestyle=dotted](-1,-1)(4,4)
\psline[linestyle=dotted](1,-1)(-4,4)
\psparametricplot[plotstyle=curve,linewidth=1.5pt]{-1.1}{1.1}{t SINH 3 mul t COSH 3 mul}
\psline[showpoints=true](0,0)(-1.232,3.243)
\rput(-1.462,3.023){$C$}
\psarc(0,0){0.6}{90}{110.8}
\rput(-0.14,0.8){$\theta$}
\psarc(0,5){0.6}{270}{290.8}
\rput(0.14,4.2){$\theta$}
\psline{->}(-4,4.294)(4,1.255)
\rput(4.45,1.255){$i c t'$}
\psline[showpoints=true](0,2.775)(0,5)
\rput(-0.25,2.575){$U$}
\rput(0.25,5.25){$S$}
\pspolygon*(0,5)(-1.04,5.46)(-0.96,5.54)          
\rput(-1.2,5.7){$A$}
\pspolygon*(0.9877,2.3998)(-1.04,5.46)(-0.96,5.54)          
\rput(1.2377,2.5998){$T$}
\psline(0,5)(0.9877,2.3998)
\psline[linestyle=dotted](-1.232,3.243)(-1,5.5)
\psdot(-1,5.5)     
\psdot(0.9877,2.3998)      
\end{pspicture}
\caption{In the reference frame $K'$ where the retarded source particle at $C$
is instantaneously at rest, the position of the field point $A$ is given by $\overrightarrow{TA}$.}
\label{fig:fig8}
\end{center}
\end{figure}

Through point $C$ we draw the line tangent to the hyperbola, 
which intersects the $x$ axis at point $U$, as seen in Fig. \ref{fig:fig8}.
This tangent line is the $i c t'$ time axis of 
the inertial reference frame $K'$ in which the retarded source particle at $C$ is instantaneously at rest.
We also notice that the tangent line $CU$ is perpendicular to $OC$, as confirmed by the 
equation $\mathbf{X}_1 \cdot \mathbf{V}_1= 0$.

From point $A$ we draw a line perpendicular to the $i c t'$ axis, which intersects it at point $T$.
The $i c t'$ axis is perpendicular to both $AT$ (by construction) and $AS$ (since $AS$ is perpendicular
to any line in the $x O i c t$ plane), and as a result the $i c t'$ axis is perpendicular to the $AST$ plane.
As a consequence, the $i c t'$ axis is perpendicular to $ST$.
Since the $i c t'$ axis is also perpendicular to $OC$,
we conclude that $OC$ is parallel to $ST$, and as a result the measure of $\angle UST$ is also $\theta$.

We have \cite{GaleriuHYP}
\begin{eqnarray}
OC = a, \\
OU = OC / \cos(\theta) = a / \cos(\theta), \\
OS = \rho, \\
AS = \sqrt{y^2 + z^2}, \\
US = OS - OU = \rho - a / \cos(\theta), \\
TS = US \cos(\theta) = \rho \cos(\theta) - a, \\
TA = \sqrt{TS^2 + AS^2} = \sqrt{\rho^2 \cos^2(\theta) + a^2  - 2 \rho a \cos(\theta) + y^2 + z^2}.
\end{eqnarray}
With the help of equation(\ref{eq:costheta}) we find that
\begin{equation}
TA = \sqrt{\rho^2 \cos^2(\theta) - \rho^2} = \rho \sqrt{\cos^2(\theta) - 1} = - i \rho \sin(\theta). 
\end{equation}

We introduce the notation $\overrightarrow{TA} = (\vec{r}, 0)$, 
meaning that in reference frame $K'$ we have
\begin{equation}
\vec{r} = \Big(\rho \cos(\theta) - a, y, z\Big).
\end{equation}
The length of segment $TA$ is the Coulombian radius $r$ that enters the calculation of the 
four-force in a reference frame co-moving with the source particle. 
The distance
\begin{equation}
r = - i \rho \sin(\theta)
\end{equation}
is a Lorentz invariant quantity.
The {\bf magnitude of the electric field} at point $A$,
as shown in Ref. \cite{GaleriuHYP}, is $E = Q/r^2$ (in Gaussian units).

In conclusion, in the reference frame
where the field point and the center of the hyperbola are simultaneous,
the particle in hyperbolic motion is the source of an electric field $\vec{E}$ 
with the direction of $\overrightarrow{MA}$
and the magnitude of $Q/r^2$. 
In this reference frame the magnetic field $\vec{B}$
is zero and the Lorentz force has only an electric part, that is $\vec{F} = q \, \vec{E}$.

A Lorentz transformation could bring us to a reference frame
where the field point and the center of the hyperbola are not simultaneous.
This case, which produces the general expression of the electromagnetic field tensor, will not be investigated here.

\subsection{Test particle at rest}

Consider a test particle with electric charge $q$, at rest at the position of field point $A$.
We work in the reference frame $K$ where the field point and the center of the hyperbola are simultaneous.

The four-velocity of the retarded source particle at point $C$ is
\begin{equation}
\mathbf{V}_{1ret} = \Big(i \, c \, \sin(\theta), 0, 0, i \, c \, \cos(\theta)\Big).
\label{eq:4velocityQ}
\end{equation}
The four-velocity of the advanced source particle at point $E$ is
\begin{equation}
\mathbf{V}_{1adv} = \Big(- i \, c \, \sin(\theta), 0, 0, i \, c \, \cos(\theta)\Big).
\label{eq:4velocityN}
\end{equation}
The four-velocity of the test particle at rest is
\begin{equation}
\mathbf{V}_2 = (\vec{0}, i c) = (0, 0, 0, i c).
\label{eq:rest_v2_hyp}
\end{equation}
For the retarded interaction, the displacement four-vector is
\begin{equation}
\mathbf{X}_2 - \mathbf{X}_{1 ret} = (\vec{R}, i R) = \Big(\rho - a \cos(\theta), y, z, i R\Big).
\label{eq:rest_ret_hyp}
\end{equation}
For the advanced interaction, the displacement four-vector is
\begin{equation}
\mathbf{X}_2 - \mathbf{X}_{1 adv} = (\vec{R}, - i R) = \Big(\rho - a \cos(\theta), y, z, - i R\Big).
\label{eq:rest_adv_hyp}
\end{equation}
The ratios of the corresponding segments are calculated according to formulas (\ref{eq:AB_CD_new}) and (\ref{eq:AB_EF_new}).
For the retarded interaction
\begin{multline}
\frac{AB}{CD} 
= \frac{(\mathbf{X}_2 - \mathbf{X}_{1 ret}) \cdot \mathbf{V}_{1ret}}{(\mathbf{X}_2 - \mathbf{X}_{1 ret}) \cdot \mathbf{V}_2} \\
= \frac{\Big(\rho - a \cos(\theta), y, z, i R\Big) \cdot \Big(i \, c \, \sin(\theta), 0, 0, i \, c \, \cos(\theta)\Big)}{\Big(\rho - a \cos(\theta), y, z, i R\Big) \cdot (0, 0, 0, i c)} \\
= \frac{i \, c \, \sin(\theta) [\rho - a \cos(\theta)] - R c \cos(\theta)}{- R c}
= \frac{\rho}{a},
\label{eq:galeriu1resthyp}
\end{multline}
and for the advanced interaction
\begin{multline}
\frac{AB}{EF} 
= \frac{(\mathbf{X}_2 - \mathbf{X}_{1 adv}) \cdot \mathbf{V}_{1adv}}{(\mathbf{X}_2 - \mathbf{X}_{1 adv}) \cdot \mathbf{V}_2} \\
= \frac{\Big(\rho - a \cos(\theta), y, z, - i R\Big) \cdot \Big(- i \, c \, \sin(\theta), 0, 0, i \, c \, \cos(\theta)\Big)}{\Big(\rho - a \cos(\theta), y, z, - i R\Big) \cdot (0, 0, 0, i c)} \\
= \frac{- i \, c \, \sin(\theta) [\rho - a \cos(\theta)] + R c \cos(\theta)}{R c}
= \frac{\rho}{a},
\label{eq:galeriu2resthyp}
\end{multline}
where we have used the fact that $i \sin(\theta) = - R / a$.
We also notice that $r / R = \rho / a$, in full agreement with our conclusion from Appendix B, since 
$r$ is the Coulombian radius to the test particle in the co-moving reference frame of the source particle,
while
$R$ is the Coulombian radius to the (retarded or advanced) source particle 
in the proper reference frame of the test particle.

According to the mechanism described, 
in the reference frame $K'$ that is co-moving with the
retarded source particle at $C$,
the linear four-force density acting on the particle at $A$ is
\begin{equation}
\mathbf{f}_{ret} = \frac{-1}{s_o} \Big( \frac{Q q}{2 r^2} \frac{\rho}{a} \frac{\vec{r}}{r}, i \frac{Q q}{2 r^2} \frac{\rho}{a}\Big)
= \frac{-1}{s_o} \frac{Q q}{2 r^3} \frac{\rho}{a} (\vec{r}, i r).
\label{fpret_hyp_0}
\end{equation}
What are the components of this linear four-force density in the reference frame $K$ in which 
the field point and the center of the hyperbola are simultaneous?
In reference frame $K'$ we have
\begin{equation}
\mathbf{X}_2 - \mathbf{X}_{1 ret} = \overrightarrow{CA} 
= \overrightarrow{CT} + \overrightarrow{TA} = (\vec{0}, i r) + (\vec{r}, 0) = (\vec{r}, i r),
\end{equation}
while in reference frame $K$ we have
\begin{equation}
\mathbf{X}_2 - \mathbf{X}_{1 ret} = \overrightarrow{CA} 
= \overrightarrow{CM} + \overrightarrow{MA} = (\vec{0}, i R) + (\vec{R}, 0) = (\vec{R}, i R).
\end{equation}
As a result, in reference frame $K$ we have 
\begin{equation}
\mathbf{f}_{ret} 
= \frac{-1}{s_o} \frac{Q q}{2 r^3} \frac{\rho}{a} (\vec{R}, i R)
= \frac{-1}{s_o} \Big( \frac{Q q}{2 r^2} \frac{\rho}{a} \frac{\vec{R}}{R} \frac{R}{r}, 
i \frac{Q q}{2 r^2} \frac{\rho}{a} \frac{R}{r}\Big)
= \frac{-1}{s_o} \Big( \frac{Q q}{2 r^2} \frac{\vec{R}}{R}, i \frac{Q q}{2 r^2}\Big).
\label{fret_hyp_0}
\end{equation}
In a similar manner we derive
\begin{equation}
\mathbf{f}_{adv} = \frac{-1}{s_o} \Big( \frac{Q q}{2 r^2} \frac{\vec{R}}{R}, - i \frac{Q q}{2 r^2}\Big).
\label{fadv_hyp_0}
\end{equation}
The total linear four-force density becomes
\begin{equation}
\mathbf{f} = \mathbf{f}_{ret} + \mathbf{f}_{adv} 
= \frac{-1}{s_o} \Big(\frac{Q q}{r^2} \frac{\vec{R}}{R}, 0\Big),
\label{f_hyp_0}
\end{equation}
and the total four-force $\mathbf{F} = - s_0 \, \mathbf{f}$
is identical to the classical result.

We notice that, in the proper reference frame of the test particle,
in order for the four-force to be orthogonal to the four-velocity,
we need the imaginary (temporal) components
of the retarded and advanced linear four-force densities to cancel each other. This happens
for a source charge in hyperbolic motion, in the reference frame in which 
the field point and the center of the hyperbola are simultaneous,
due to the symmetry of the configuration,
but this may not happen in general for a source charge in random motion.

\subsection{Test particle in motion}

This time the four-velocity of the test particle at the position of the field point $A$ is
\begin{equation}
\mathbf{V}_2 = (\gamma_2 \, \vec{v_2}, i \gamma_2 \, c).
\label{eq:v2_hyp}
\end{equation}

The ratios of the corresponding segments are calculated according to formulas (\ref{eq:AB_CD_new}) and (\ref{eq:AB_EF_new}).
Only the denominators in equations (\ref{eq:galeriu1resthyp})-(\ref{eq:galeriu2resthyp}) are a little bit different.
For the retarded interaction we have
\begin{multline}
(\mathbf{X}_2 - \mathbf{X}_{1 ret}) \cdot \mathbf{V}_2 = (\vec{R}, i R) \cdot (\gamma_2 \, \vec{v_2}, i \gamma_2 \, c) \\
= \gamma_2 \vec{R} \cdot \vec{v_2} - \gamma_2 R c = - R c \, \gamma_2 \, (1 - v_{2rad} / c), 
\end{multline}
\begin{equation}
\frac{AB}{CD} 
= \frac{(\mathbf{X}_2 - \mathbf{X}_{1 ret}) \cdot \mathbf{V}_{1ret}}{(\mathbf{X}_2 - \mathbf{X}_{1 ret}) \cdot \mathbf{V}_2}
= \frac{\rho}{a} \frac{1}{\gamma_2 \, (1 - v_{2rad} / c)},
\label{eq:galeriu1hyp}
\end{equation}
and for the advanced interaction
\begin{multline}
(\mathbf{X}_2 - \mathbf{X}_{1 adv}) \cdot \mathbf{V}_2 = (\vec{R}, - i R) \cdot (\gamma_2 \, \vec{v_2}, i \gamma_2 \, c) \\
= \gamma_2 \vec{R} \cdot \vec{v_2} + \gamma_2 R c = R c \, \gamma_2 \, (1 + v_{2rad} / c), 
\end{multline}
\begin{equation}
\frac{AB}{EF} 
= \frac{(\mathbf{X}_2 - \mathbf{X}_{1 adv}) \cdot \mathbf{V}_{1adv}}{(\mathbf{X}_2 - \mathbf{X}_{1 adv}) \cdot \mathbf{V}_2}
= \frac{\rho}{a} \frac{1}{\gamma_2 \, (1 + v_{2rad} / c)},
\label{eq:galeriu2hyp}
\end{equation}
where by definition $v_{2rad} = \vec{v_2} \cdot \vec{R} / R$.

In reference frame $K$ the retarded linear four-force density becomes
\begin{equation}
\mathbf{f}_{ret} = \frac{-1}{s_o} \Big( \frac{Q q}{2 r^2} \frac{1}{\gamma_2 \, (1 - v_{2rad} / c)} \frac{\vec{R}}{R}, 
i \frac{Q q}{2 r^2} \frac{1}{\gamma_2 \, (1 - v_{2rad} / c)}\Big),
\label{fret_hyp}
\end{equation}
and the advanced linear four-force density becomes
\begin{equation}
\mathbf{f}_{adv} = \frac{-1}{s_o} \Big( \frac{Q q}{2 r^2} \frac{1}{\gamma_2 \, (1 + v_{2rad} / c)} \frac{\vec{R}}{R}, 
- i \frac{Q q}{2 r^2} \frac{1}{\gamma_2 \, (1 + v_{2rad} / c)}\Big).
\label{fadv_hyp}
\end{equation}
The total linear four-force density becomes
\begin{equation}
\mathbf{f} = \mathbf{f}_{ret} + \mathbf{f}_{adv} 
= \frac{-1}{s_o} \Big(\gamma_2 \frac{Q q}{r^2} \frac{\vec{R}}{R}, i \gamma_2 \frac{v_{2rad}}{c} \frac{Q q}{r^2}\Big) 
\frac{1}{\gamma_2^{\ 2} \, (1 - v_{2rad}^{\ 2} / c^2)}.
\label{f_hyp}
\end{equation}
The total four-force $\mathbf{F} = - s_0 \, \mathbf{f}$ is perpendicular to the four-velocity of the test particle,
as required by the first part of equation (\ref{eq:fourforce}).

We again notice that there is an extra factor
\begin{equation}
\Gamma_2 = \frac{1}{\gamma_2^{\ 2} \, (1 - v_{2rad}^{\ 2} / c^2)} = \frac{c^2 - v_2^{\ 2}}{c^2 - v_{2rad}^{\ 2}}
\end{equation}
that makes our expression of the four-force a little bit different from what we expect from classical electrodynamics.

\section{The invariant expression of the four-force}

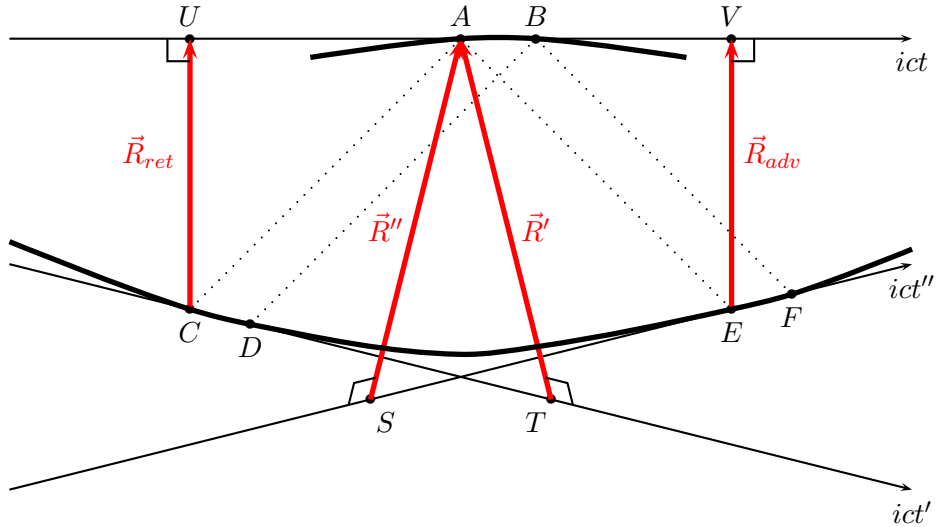
\begin{figure}[h!]
\begin{center}
\begin{pspicture}(-6,0.5)(6,-6)
\psline{->}(-6,0)(6,0)
\rput(6,-0.3){$i c t$}
\psdot(0,0)
\rput(0,0.3){$A$}
\psdot(1,0)
\rput(1,0.3){$B$}
\psline{->}(-6,-3)(6,-6)
\rput(6,-6.3){$i c t'$}
\psdot(-3.6,-3.6)
\rput(-3.6,-3.9){$C$}
\psdot(-2.8,-3.8)
\rput(-2.8,-4.1){$D$}
\psline{->}(-6,-6)(6,-3)
\rput(6,-3.3){$i c t''$}
\psdot(3.6,-3.6)
\rput(3.6,-3.9){$E$}
\psdot(4.4,-3.4)
\rput(4.4,-3.7){$F$}
\psdot(-1.2,-4.8)
\rput(-1.0,-5.1){$S$}
\psdot(1.2,-4.8)
\rput(1.0,-5.1){$T$}
\psdot(-3.6,0)
\rput(-3.6,0.3){$U$}
\psdot(3.6,0)
\rput(3.6,0.3){$V$}
\psline[linestyle=dotted](-3.6,-3.6)(0,0)   
\psline[linestyle=dotted](-2.8,-3.8)(1,0)   
\psline[linestyle=dotted](0,0)(3.6,-3.6)   
\psline[linestyle=dotted](1,0)(4.4,-3.4)   
\psline[linecolor=red,linewidth=2pt]{->}(-1.2,-4.8)(0,0)   
\psline[linecolor=red,linewidth=2pt]{->}(1.2,-4.8)(0,0)   
\psline[linecolor=red,linewidth=2pt]{->}(-3.6,-3.6)(-3.6,0)   
\psline[linecolor=red,linewidth=2pt]{->}(3.6,-3.6)(3.6,0)   
\psline(-3.9,0)(-3.9,-0.3)(-3.6,-0.3)   
\psline(3.9,0)(3.9,-0.3)(3.6,-0.3)   
\psline(-1.12724,-4.50896)(-1.41828,-4.58172)(-1.49104,-4.87276)   
\psline(1.12724,-4.50896)(1.41828,-4.58172)(1.49104,-4.87276)   
\pscurve[linewidth=2pt](-2,-0.25)(0,0)(1,0)(3,-0.25)
\pscurve[linewidth=2pt](-6,-2.7)(-3.6,-3.6)(-2.8,-3.8)(0,-4.2)(1,-4.1)(3.6,-3.6)(4.4,-3.4)(6,-2.8)
\rput(1,-2.5){\textcolor{red}{$\vec{R}'$}}
\rput(-1,-2.5){\textcolor{red}{$\vec{R}''$}}
\rput(-4.15,-1.5){\textcolor{red}{$\vec{R}_{ret}$}}
\rput(4.15,-1.5){\textcolor{red}{$\vec{R}_{adv}$}}
\end{pspicture}
\caption{The test particle at $AB$ interacts with the retarded source particle at $CD$
and with the advanced source particle at $EF$. The time axes are lines tangent to the worldlines
of the particles through points $A$, $C$, and $E$.}
\label{fig:9}
\end{center}
\end{figure}

We are now ready to write down the general expression of the four-force
in the reference frame $K$ that is co-moving with the test particle at point $A$.
Since we will write this expression in an explicitly Lorentz invariant form,
the expression of the four-force will be equally valid in any inertial reference frame.
Let $K'$ be the reference frame that is co-moving with the retarded source particle at point $C$,
and let $K''$ be the reference frame that is co-moving with the advanced source particle at point $E$.
As seen from Figure \ref{fig:9}, 
the $i c t$ time axis is the line that goes through point $A$ and is tangent to the worldline of the test particle,
the $i c t'$ time axis is the line that goes through point $C$ and is tangent to the worldline of the retarded source particle,
and the $i c t''$ time axis is the line that goes through point $E$ and is tangent to the worldline of the advanced source particle.

From point $A$ we draw a line perpendicular to the $i c t'$ axis, which intersects it at point $T$. 
The length of segment $TA$ is $R'$, and the length of segment $CT$ is $i R'$.
In reference frame $K'$ we have $\overrightarrow{TA} = (\vec{R}', 0)$.
From point $C$ we draw a line perpendicular to the $i c t$ axis, which intersects it at point $U$. 
The length of segment $UC$ is $R_{ret}$, and the length of segment $UA$ is $i R_{ret}$.
In reference frame $K$ we have $\overrightarrow{CU} = (\vec{R}_{ret}, 0)$.

From point $A$ we draw a line perpendicular to the $i c t''$ axis, which intersects it at point $S$. 
The length of segment $SA$ is $R''$, and the length of segment $SE$ is $i R''$.
In reference frame $K''$ we have $\overrightarrow{SA} = (\vec{R}'', 0)$.
From point $E$ we draw a line perpendicular to the $i c t$ axis, which intersects it at point $V$. 
The length of segment $EV$ is $R_{adv}$, and the length of segment $AV$ is $i R_{adv}$.
In reference frame $K$ we have $\overrightarrow{EV} = (\vec{R}_{adv}, 0)$.

In reference frame $K'$ the retarded linear four-force density is
\begin{equation}
\mathbf{f}_{ret} = \frac{-1}{s_o} \Big( \frac{Q q}{2 R'^2} \frac{\vec{R}'}{R'} \frac{AB}{CD}, 
i \frac{Q q}{2 R'^2} \frac{AB}{CD}\Big)
= \frac{-1}{s_o} \frac{Q q}{2 R'^3} \frac{AB}{CD} (\vec{R}', i R'),
\label{eq97}
\end{equation}
and the displacement four-vector $\overrightarrow{CA}$ is
\begin{equation}
\mathbf{X}_2 - \mathbf{X}_{1ret} = \overrightarrow{CA} = \overrightarrow{CT} + \overrightarrow{TA} 
= (\vec{0}, i R') + (\vec{R}', 0) = (\vec{R}', i R').
\end{equation}
In reference frame $K$ the displacement four-vector $\overrightarrow{CA}$ is
\begin{equation}
\mathbf{X}_2 - \mathbf{X}_{1ret} = \overrightarrow{CA} = \overrightarrow{CU} + \overrightarrow{UA} 
= (\vec{R}_{ret}, 0) + (\vec{0}, i R_{ret}) = (\vec{R}_{ret}, i R_{ret}),
\end{equation}
and, as a consequence, the retarded linear four-force density is
\begin{equation}
\mathbf{f}_{ret} = \frac{-1}{s_o} \frac{Q q}{2 R'^3} \frac{AB}{CD} (\vec{R}_{ret}, i R_{ret})
= \frac{-1}{s_o} \Big( \frac{Q q}{2 R'^2} \frac{\vec{R}_{ret}}{R_{ret}} \frac{AB}{CD} \frac{R_{ret}}{R'}, 
i \frac{Q q}{2 R'^2} \frac{AB}{CD} \frac{R_{ret}}{R'}\Big),
\label{eq98}
\end{equation}
which, since $R' / R_{ret} = AB / CD$ according to formula (\ref{eq:ratio2}), simplifies to
\begin{equation}
\mathbf{f}_{ret} = \frac{-1}{s_o} \Big( \frac{Q q}{2 R'^2} \frac{\vec{R}_{ret}}{R_{ret}}, i \frac{Q q}{2 R'^2}\Big).
\end{equation}

In reference frame $K'$, where $\mathbf{V}_{1ret} = (\vec{0}, i c)$, we have
\begin{equation}
(\mathbf{X}_2 - \mathbf{X}_{1ret}) \cdot \mathbf{V}_{1ret} = (\vec{R}', i R') \cdot (\vec{0}, i c) = - R' \, c,
\end{equation}
and in reference frame $K$, where $\mathbf{V}_2 = (\vec{0}, i c)$, we have
\begin{equation}
(\mathbf{X}_2 - \mathbf{X}_{1ret}) \cdot \mathbf{V}_2 = (\vec{R}_{ret}, i R_{ret}) \cdot (\vec{0}, i c) = - R_{ret} \, c.
\end{equation}
The Lorentz invariant expression of the retarded four-force becomes
\begin{multline}
\mathbf{F}_{ret} = - s_0 \, \mathbf{f}_{ret}
= \frac{Q \, q}{2 \, R'^2 \, R_{ret}} (\vec{R}_{ret}, i R_{ret}) \\
= \frac{- Q \, q \, c^3 \, (\mathbf{X}_2 - \mathbf{X}_{1ret})}{2 \, [(\mathbf{X}_2 - \mathbf{X}_{1ret}) \cdot \mathbf{V}_{1ret}]^2 \, (\mathbf{X}_2 - \mathbf{X}_{1ret}) \cdot \mathbf{V}_2}. 
\label{eq99}
\end{multline}

In reference frame $K''$ the advanced linear four-force density is
\begin{equation}
\mathbf{f}_{adv} = \frac{-1}{s_o} \Big( \frac{Q q}{2 R''^2} \frac{\vec{R}''}{R''} \frac{AB}{EF}, 
- i \frac{Q q}{2 R''^2} \frac{AB}{EF}\Big)
= \frac{-1}{s_o} \frac{Q q}{2 R''^3} \frac{AB}{EF} (\vec{R}'', - i R''),
\label{eq100}
\end{equation}
and the displacement four-vector $\overrightarrow{EA}$ is
\begin{equation}
\mathbf{X}_2 - \mathbf{X}_{1adv} = \overrightarrow{EA} = \overrightarrow{ES} + \overrightarrow{SA} 
= (\vec{0}, - i R'') + (\vec{R}'', 0) = (\vec{R}'', - i R'').
\end{equation}
In reference frame $K$ the displacement four-vector $\overrightarrow{EA}$ is
\begin{equation}
\mathbf{X}_2 - \mathbf{X}_{1adv} = \overrightarrow{EA} = \overrightarrow{EV} + \overrightarrow{VA} 
= (\vec{R}_{adv}, 0) + (\vec{0}, - i R_{adv}) = (\vec{R}_{adv}, - i R_{adv}),
\end{equation}
and, as a consequence, the advanced linear four-force density is
\begin{multline}
\mathbf{f}_{adv} = \frac{-1}{s_o} \frac{Q q}{2 R''^3} \frac{AB}{EF} (\vec{R}_{adv}, - i R_{adv}) \\
= \frac{-1}{s_o} \Big( \frac{Q q}{2 R''^2} \frac{\vec{R}_{adv}}{R_{adv}} \frac{AB}{EF} \frac{R_{adv}}{R''}, 
- i \frac{Q q}{2 R''^2} \frac{AB}{EF} \frac{R_{adv}}{R''}\Big),
\label{eq101}
\end{multline}
which, since $R'' / R_{adv} = AB / EF$ according to formula (\ref{eq:ratio2}), simplifies to
\begin{equation}
\mathbf{f}_{adv} = \frac{-1}{s_o} \Big( \frac{Q q}{2 R''^2} \frac{\vec{R}_{adv}}{R_{adv}}, - i \frac{Q q}{2 R''^2}\Big).
\end{equation}

In reference frame $K''$, where $\mathbf{V}_{1adv} = (\vec{0}, i c)$, we have
\begin{equation}
(\mathbf{X}_2 - \mathbf{X}_{1adv}) \cdot \mathbf{V}_{1adv} = (\vec{R}'', - i R'') \cdot (\vec{0}, i c) = R'' \, c,
\end{equation}
and in reference frame $K$, where $\mathbf{V}_2 = (\vec{0}, i c)$, we have
\begin{equation}
(\mathbf{X}_2 - \mathbf{X}_{1adv}) \cdot \mathbf{V}_2 = (\vec{R}_{adv}, - i R_{adv}) \cdot (\vec{0}, i c) = R_{adv} \, c.
\end{equation}
The Lorentz invariant expression of the advanced four-force becomes
\begin{multline}
\mathbf{F}_{adv} = - s_0 \, \mathbf{f}_{adv}
= \frac{Q \, q}{2 \, R''^2 \, R_{adv}} (\vec{R}_{adv}, - i R_{adv}) \\
= \frac{Q \, q \, c^3 \, (\mathbf{X}_2 - \mathbf{X}_{1adv})}{2 \, [(\mathbf{X}_2 - \mathbf{X}_{1adv}) \cdot \mathbf{V}_{1adv}]^2 \, (\mathbf{X}_2 - \mathbf{X}_{1adv}) \cdot \mathbf{V}_2}. 
\label{eq102}
\end{multline}

Adding the two contributions we obtain the total four-force
\begin{multline}
\mathbf{F} = \mathbf{F}_{ret} + \mathbf{F}_{adv} 
= \frac{Q \, q}{2 \, R'^2 \, R_{ret}} (\vec{R}_{ret}, i R_{ret}) + \frac{Q \, q}{2 \, R''^2 \, R_{adv}} (\vec{R}_{adv}, - i R_{adv}) \\
= \frac{- Q \, q \, c^3 \, (\mathbf{X}_2 - \mathbf{X}_{1ret})}{2 \, [(\mathbf{X}_2 - \mathbf{X}_{1ret}) \cdot \mathbf{V}_{1ret}]^2 \, (\mathbf{X}_2 - \mathbf{X}_{1ret}) \cdot \mathbf{V}_2} \\
+ \frac{Q \, q \, c^3 \, (\mathbf{X}_2 - \mathbf{X}_{1adv})}{2 \, [(\mathbf{X}_2 - \mathbf{X}_{1adv}) \cdot \mathbf{V}_{1adv}]^2 \, (\mathbf{X}_2 - \mathbf{X}_{1adv}) \cdot \mathbf{V}_2}.
\label{total_invariant_4F}
\end{multline}

Our expression (\ref{total_invariant_4F}) of the total four-force is closely related to the electrostatic (\lq\lq elektrostatischen\rq\rq)
four-force of Fokker \cite{fokker}, which is \cite{arcane}
\begin{multline}
\mathbf{F} = \mathbf{F}_{ret} + \mathbf{F}_{adv} = 
\frac{- Q \, q \, (\mathbf{V}_2 \cdot \mathbf{V}_{1ret})^2 \, (\mathbf{X}_2 - \mathbf{X}_{1ret})}{2 \, c \, [(\mathbf{X}_2 - \mathbf{X}_{1ret}) \cdot \mathbf{V}_{1ret}]^2 \, (\mathbf{X}_2 - \mathbf{X}_{1ret}) \cdot \mathbf{V}_2} \\
+ \frac{Q \, q \, (\mathbf{V}_2 \cdot \mathbf{V}_{1adv})^2 \, (\mathbf{X}_2 - \mathbf{X}_{1adv})}{2 \, c \, [(\mathbf{X}_2 - \mathbf{X}_{1adv}) \cdot \mathbf{V}_{1adv}]^2 \, (\mathbf{X}_2 - \mathbf{X}_{1adv}) \cdot \mathbf{V}_2}.
\label{Fokker_electrostatic_4F}
\end{multline}
We can get formula (\ref{total_invariant_4F}) from formula (\ref{Fokker_electrostatic_4F}) if 
we replace $(\mathbf{V}_2 \cdot \mathbf{V}_1)^2$ with 
$\mathbf{V}_2^{\ 2} \, \mathbf{V}_1^{\ 2} = c^4$.
For this reason it would be interesting to see what happens when the scalar product $\mathbf{V}_2 \cdot \mathbf{V}_1$
in Fokker's electrodynamic action \cite{fokker, Wheeler1949, galeriu2023} is replaced with 
$| \mathbf{V}_2 | \, | \mathbf{V}_1 |$. 
Since $\mathbf{V}_2 \cdot \mathbf{V}_1 = | \mathbf{V}_2 | \, | \mathbf{V}_1 | \, \cos(\varphi)$, 
where $\varphi$ is the angle between the
four-velocities, we are justified in making this substitution whenever the two four-velocities are parallel,
or whenever the relative velocities of the two particles are very small. This condition is also assumed true
in Synge's theory \cite{synge, Synge1935}. 
With this substitution Fokker's action, written with a metric tensor of signature $(+, -, -, -)$
\begin{multline}
W_{Fokker} = - \sum_A \int  m_A c \sqrt{dx_{A \alpha} \, dx_A^\alpha} \\
 - \sum_A \sum_{B > A} \frac{q_A q_B}{c} \int \int 
\delta\big( (x_{A \beta} - x_{B \beta}) (x_A^\beta - x_B^\beta) \big) \, dx_{A \alpha} \, dx_B^\alpha,
\label{eq:actionFokker}
\end{multline}
becomes
\begin{multline}
W = - \sum_A \int  m_A c \sqrt{dx_{A \alpha} \, dx_A^\alpha} \\
 - \sum_A \sum_{B > A} \frac{q_A q_B}{c} \int \int 
\delta\big( (x_{A \beta} - x_{B \beta}) (x_A^\beta - x_B^\beta) \big) \, 
\sqrt{dx_{A \mu} \, dx_A^\mu} \, \sqrt{dx_{B \nu} \, dx_B^\nu}.
\label{eq:actionGaleriu}
\end{multline}

\section{The principle of action and reaction}

The principle of action and reaction, 
which has already been verified for two particles with relative motion only in the radial direction,
will now be demonstrated for the most general situation.
We work in reference frame $K$.

The retarded four-force with which particle $CD$ acts on particle $AB$ is
\begin{equation}
\mathbf{f}_{ret} \times AB = \frac{-1}{s_o} \Big( \frac{Q q}{2 R'^2} \frac{\vec{R}_{ret}}{R_{ret}} AB, 
i \frac{Q q}{2 R'^2} AB\Big).
\end{equation}
The advanced four-force with which particle $AB$ acts on particle $CD$ is
\begin{equation}
\mathbf{g}_{adv} \times CD = \frac{-1}{s_o} \Big(- \frac{Q q}{2 R_{ret}^2} \frac{\vec{R}_{ret}}{R_{ret}} \frac{CD}{AB} CD, 
- i \frac{Q q}{2 R_{ret}^2} \frac{CD}{AB} CD\Big).
\end{equation}
We notice that $\mathbf{g}_{adv} \times CD = - \mathbf{f}_{ret} \times AB$,
since $\dfrac{1}{R'^2} AB = \dfrac{1}{R_{ret}^2} \dfrac{CD}{AB} CD$.

The advanced four-force with which particle $EF$ acts on particle $AB$ is
\begin{equation}
\mathbf{f}_{adv} \times AB = \frac{-1}{s_o} \Big( \frac{Q q}{2 R''^2} \frac{\vec{R}_{adv}}{R_{adv}} AB, 
- i \frac{Q q}{2 R''^2} AB\Big).
\end{equation}
The retarded four-force with which particle $AB$ acts on particle $EF$ is
\begin{equation}
\mathbf{g}_{ret} \times EF = \frac{-1}{s_o} \Big(- \frac{Q q}{2 R_{adv}^2} \frac{\vec{R}_{adv}}{R_{adv}} \frac{EF}{AB} EF, 
i \frac{Q q}{2 R_{adv}^2} \frac{EF}{AB} EF\Big).
\end{equation}
We notice that $\mathbf{g}_{ret} \times EF = - \mathbf{f}_{adv} \times AB$,
since $\dfrac{1}{R''^2} AB = \dfrac{1}{R_{adv}^2} \dfrac{EF}{AB} EF$.

When going from the worldline string model back to the material point particle model, 
the product of the linear four-force density with the length of the infinitesimal worldline segment
is replaced by the product of the four-force with the infinitesimal change in proper time,
an expression equal to the change in four-momentum. 
The principle of action and reaction that we have in the worldline string model 
is equivalent to 
the law of conservation of total four-momentum that we have in the material point particle model.

\section{Concluding remarks}

This is the third manuscript in which we have investigated a time symmetric action-at-a-distance theory 
of electrodynamic interaction. Some of the assumptions put forward in the first manuscript \cite{galeriuAAAD}
have been proven wrong, and they have been replaced with different assumptions in the second manuscript \cite{arcane}. 
In the final formulation our theory is based on two postulates.

Postulate I. The interaction is time symmetric,
with the retarded and the advanced parts on equal footing. 
The retarded and the advanced four-forces (the exchanges in four-momentum) 
are parallel to the displacement four-vectors connecting the two interacting particles.

This means that we implement \lq\lq the parallel condition\rq\rq. 

Postulate II. The interaction takes place
between worldline segments of infinitesimal length, whose end points are connected by light signals.
In the inertial reference frame in which the source particle is instantaneously at rest, 
the linear four-force density acting on the test particle
is proportional to the product of the Coulombian electrostatic force with the ratio of 
the length of the infinitesimal segment on the worldline of the test particle 
to the length of the corresponding infinitesimal segment on the worldline of the source particle.

This is how we take into consideration \lq\lq the thickness of the light-cone\rq\rq. 

In the present work some new geometrical and algebraic derivations are presented, 
with the goal of making the overall exposition clearer.
The action and reaction principle, proven to work in a particular situation in \cite{arcane}, is here demonstrated in the general case.
We have also derived the Lorentz invariant expression of our electrodynamic four-force.
This proposed four-force (\ref{total_invariant_4F}) matches with a very good approximation, 
or even exactly (when the relative velocity has the radial direction), the classical expression of the electrodynamic four-force.

We also notice that, in our theory, when the source particle is in uniform or in hyperbolic motion,
the four-force acting on the test particle is orthogonal to the four-velocity of the test particle.
This orthogonality does not necessarily happen for a source charge in random motion.
Thus, in our theory, the rest mass of a particle will change a little bit during interactions.
The variation of the rest mass is something that also shows up in special conformal transformations.
For a very simple example we can prove that the rest mass is the same before and after the interaction \cite{galeriuAAAD}.
We hope that this result holds in general.

The remarkable fact about our electrodynamic four-force (\ref{total_invariant_4F})
is that it does not depend on the acceleration of the source particle.
However, the classical expression of the four-force \cite{GaleriuHYP} shows an explicit dependence on the 
retarded four-acceleration of the source particle. 
How can we reconcile these results?
We have to remember that the classical calculation is based on the retarded Li\'{e}nard-Wiechert electromagnetic four-potential.
For a source particle at rest, or in uniform motion, or in hyperbolic motion, the advanced four-potential is equal to the 
retarded four-potential. By assuming that the higher derivatives of the retarded position four-vector of the source particle
(the derivative of the four-acceleration, etc.)
make no contribution to the electrodynamic four-force, 
we are able to replace the
advanced part of the electrodynamic four-potential with the corresponding retarded part.
This is equivalent to the writing of the advanced position
four-vector and of the advanced four-velocity of the source particle
in the electrodynamic four-force as truncated Taylor series expansions
around the retarded position four-vector of the source particle. 
What would we have to do if the higher derivatives of the retarded position four-vector of the source particle could not be ignored?
In such a situation we would have to keep more terms in the Taylor series expansions of
the advanced position four-vector and of the advanced four-velocity of the source particle 
around the retarded position four-vector of the source particle, thus bringing into
the expression of the electrodynamic four-force not only the retarded four-acceleration of the source particle, 
but also the higher derivatives.
These extra terms, however, have negligible contributions in all practical experimental situations investigated 
in classical electrodynamics.

\section*{Appendix A. Ratio of corresponding segments} 

Consider two points connected by a light signal, 
point $C$ on the worldline of particle 1, with position four-vector $\mathbf{X}_1$,
and point $A$ on the worldline of particle 2, with position four-vector $\mathbf{X}_2$. 
Consider two other points, also connected by a light signal, and infinitely close to the first two points,
point $D$ on the worldline of particle 1, with position four-vector $\mathbf{X}_1 + d \mathbf{X}_1$,
and point $B$ on the worldline of particle 2, with position four-vector $\mathbf{X}_2 + d \mathbf{X}_2$. 
We know that 
\begin{equation}
\overrightarrow{AC} \cdot \overrightarrow{AC}
= (\mathbf{X}_1 - \mathbf{X}_2) \cdot (\mathbf{X}_1 - \mathbf{X}_2) = 0,
\label{eq:A112}
\end{equation}
and that
\begin{equation}
\overrightarrow{BD} \cdot \overrightarrow{BD}
= (\mathbf{X}_1 + d \mathbf{X}_1 - \mathbf{X}_2 - d \mathbf{X}_2) 
\cdot (\mathbf{X}_1 + d \mathbf{X}_1 - \mathbf{X}_2 - d \mathbf{X}_2) = 0.
\end{equation}
By rearanging terms in the last equation, we have
\begin{equation} 
\big((\mathbf{X}_1 - \mathbf{X}_2) + (d \mathbf{X}_1 - d \mathbf{X}_2)\big) \cdot 
\big((\mathbf{X}_1 - \mathbf{X}_2) + (d \mathbf{X}_1 - d \mathbf{X}_2)\big) = 0.
\end{equation}
By expanding the scalar product, we obtain
\begin{equation}
(\mathbf{X}_1 - \mathbf{X}_2) \cdot (\mathbf{X}_1 - \mathbf{X}_2) 
+ 2 (\mathbf{X}_1 - \mathbf{X}_2) \cdot (d \mathbf{X}_1 - d \mathbf{X}_2)
+ (d \mathbf{X}_1 - d \mathbf{X}_2) \cdot (d \mathbf{X}_1 - d \mathbf{X}_2) = 0.
\end{equation}
Due to equation (\ref{eq:A112}), to first order in the infinitesimals we have
\begin{equation}
(\mathbf{X}_1 - \mathbf{X}_2) \cdot (d \mathbf{X}_1 - d \mathbf{X}_2) = 0,
\end{equation}
an equation that we can also write as
\begin{equation}
(\mathbf{X}_1 - \mathbf{X}_2) \cdot d \mathbf{X}_1 = (\mathbf{X}_1 - \mathbf{X}_2) \cdot d \mathbf{X}_2.
\label{eq:fokker}
\end{equation}

With the substitutions $d \mathbf{X}_1 = \mathbf{V}_1 \, d \tau_1$ and $d \mathbf{X}_2 = \mathbf{V}_2 \, d \tau_2$, 
where $\mathbf{V}_1$ is the four-velocity of particle 1 at point $C$
and $\mathbf{V}_2$ is the four-velocity of particle 2 at point $A$,
the ratio of the two corresponding infinitesimal segments, connected by light signals at both ends, becomes \cite{lamont}
\begin{equation}
\frac{AB}{CD}
= \frac{i \, c \, d \tau_2}{i \, c \, d \tau_1} 
= \frac{(\mathbf{X}_1 - \mathbf{X}_2) \cdot \mathbf{V}_1}{(\mathbf{X}_1 - \mathbf{X}_2) \cdot \mathbf{V}_2}
= \frac{(\mathbf{X}_2 - \mathbf{X}_1) \cdot \mathbf{V}_1}{(\mathbf{X}_2 - \mathbf{X}_1) \cdot \mathbf{V}_2}.
\label{eq:lamont}
\end{equation}
We notice that this formula applies regardless of the temporal ordering of the two corresponding segments. 

\section*{Appendix B. Ratio of Coulombian radii}

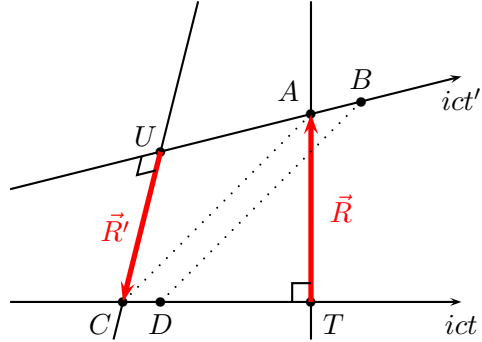
\begin{figure}[h!]
\begin{center}
\begin{pspicture}(6,-0.5)(12,4)
\psline{->}(6,0)(12,0)
\rput(12,-0.3){$i c t$}
\psline{->}(6,1.5)(12,3)
\rput(12,2.7){$i c t'$}
\psline(7.375,-0.5)(8.5,4)
\psdot(7.5,0)
\rput(7.2,-0.3){$C$}
\psdot(8,2)
\rput(7.8,2.2){$U$}
\psline[linecolor=red,linewidth=2pt]{->}(8,2)(7.5,0)
\rput(7.4,1){\textcolor{red}{$\vec{R'}$}}
\psline(10,-0.5)(10,4)
\psdot(10,0)
\rput(10.3,-0.3){$T$}
\psdot(10,2.5)
\rput(9.7,2.8){$A$}
\psline[linecolor=red,linewidth=2pt]{->}(10,0)(10,2.5)
\rput(10.4,1.25){\textcolor{red}{$\vec{R}$}}
\psline[linestyle=dotted](7.5,0)(10,2.5)        
\psline[linestyle=dotted](8,0)(10.6667,2.6667)      
\psdot(8,0)
\rput(8,-0.3){$D$}
\psdot(10.6667,2.6667)
\rput(10.6667,2.9667){$B$}
\psline(9.75,0)(9.75,0.25)(10,0.25)
\psline(7.75,1.9375)(7.6875,1.6875)(7.9375,1.75)
\end{pspicture}
\caption{Even for corresponding segments $AB$ and $CD$ that are not in the same Minkowski plane,
the relationship $AB/CD = R/R'$ still holds.}
\label{fig:figA}
\end{center}
\end{figure}

We consider the two particles from Appendix A, and we assume
that segment $CD$ is in the past of segment $AB$.
Let $i c t$ be the time axis of an inertial reference frame co-moving with the source particle at $C$,
and let $i c t'$ be the time axis of an inertial reference frame co-moving with the test particle at $A$.
From point $A$ we draw a line perpendicular to the $i c t$ axis, which intersects it at point $T$. 
The length of segment $TA$ is $R$, and the length of segment $CT$ is $i R$.
From point $C$ we draw a line perpendicular to the $i c t'$ axis, which intersects it at point $U$. 
The length of segment $UC$ is $R'$, and the length of segment $UA$ is $i R'$.
This situation is represented graphically in Fig. \ref{fig:figA}.

Let $K$ be an inertial reference frame in which particle $CD$ is instantaneously at rest at the origin,
defined up to a rotation in its 3D Euclidean space. In this reference frame
\begin{eqnarray}
\mathbf{X}_2 - \mathbf{X}_1 = \overrightarrow{CA} = \overrightarrow{CT} + \overrightarrow{TA} 
= (\vec{0}, i R) + (\vec{R}, 0) = (\vec{R}, i R), \\
\mathbf{V}_1 = (\vec{0}, i c), \\
\mathbf{V}_2 = (\gamma_2 \, \vec{v_2}, i \gamma_2 \, c), \\
\end{eqnarray}
where $\gamma_2 = \gamma(v_2) = 1 / \sqrt{1 - v_2^{\ 2} / c^2}$. We calculate the scalar products
\begin{eqnarray}
(\mathbf{X}_2 - \mathbf{X}_1) \cdot \mathbf{V}_1 = (\vec{R}, i R) \cdot (\vec{0}, i c) = - R c, \\
(\mathbf{X}_2 - \mathbf{X}_1) \cdot \mathbf{V}_2 = (\vec{R}, i R) \cdot (\gamma_2 \, \vec{v_2}, i \gamma_2 \, c) = \gamma_2 \, \vec{v_2} \cdot \vec{R} - \gamma_2 \, R c, \\
\mathbf{V}_1 \cdot \mathbf{V}_2 = (\vec{0}, i c) \cdot (\gamma_2 \, \vec{v_2}, i \gamma_2 \, c) = - \gamma_2 \, c^2.
\end{eqnarray}

Let $K'$ be an inertial reference frame in which particle $AB$ is instantaneously at rest at the origin,
defined up to a rotation in its 3D Euclidean space. In this reference frame
\begin{eqnarray}
\mathbf{X}_2 - \mathbf{X}_1 = \overrightarrow{CA} = \overrightarrow{CU} + \overrightarrow{UA} 
= (- \vec{R'}, 0) + (\vec{0}, i R') = (- \vec{R'}, i R'), \\
\mathbf{V}_1 = (\gamma'_1 \, \vec{v'_1}, i \gamma'_1 \, c), \\
\mathbf{V}_2 = (\vec{0}, i c), \\
\end{eqnarray}
where $\gamma'_1 = \gamma(v'_1) = 1 / \sqrt{1 - {v'_1}^{\, 2} / c^2}$. We calculate the scalar products
\begin{eqnarray}
(\mathbf{X}_2 - \mathbf{X}_1) \cdot \mathbf{V}_1 = (- \vec{R'}, i R') \cdot (\gamma'_1 \, \vec{v'_1}, i \gamma'_1 \, c) = - \gamma'_1 \vec{v'_1} \cdot \vec{R'} - \gamma'_1 R' c, \\
(\mathbf{X}_2 - \mathbf{X}_1) \cdot \mathbf{V}_2 = (- \vec{R'}, i R') \cdot (\vec{0}, i c) = - R' c, \\
\mathbf{V}_1 \cdot \mathbf{V}_2 = (\gamma'_1 \, \vec{v'_1}, i \gamma'_1 \, c) \cdot (\vec{0}, i c) = - \gamma'_1 c^2.
\end{eqnarray}

Since the scalar products are Lorentz invariant,
it follows that, in any reference frame, we have
\begin{equation}
\frac{R}{R'} = \frac{(\mathbf{X}_2 - \mathbf{X}_1) \cdot \mathbf{V}_1}{(\mathbf{X}_2 - \mathbf{X}_1) \cdot \mathbf{V}_2}.
\label{eq:ratio}
\end{equation}
Together with equation (\ref{eq:lamont}), we conclude that
\begin{equation}
\frac{AB}{CD} = \frac{R}{R'}.
\label{eq:ratio2}
\end{equation}

When segment $CD$ is in the future of segment $AB$, the segment is renamed $EF$. In this case
$\mathbf{X}_2 - \mathbf{X}_1 = (\vec{R}, - i R)$ in reference frame $K$,
$\mathbf{X}_2 - \mathbf{X}_1 = (- \vec{R'}, - i R')$ in reference frame $K'$,
and the same formula (\ref{eq:ratio}) for the ratio of Coulombian radii is derived. 
Formula (\ref{eq:ratio}) is left invariant by a time reversal operation 
($\mathbf{V}_1 \rightarrow - \mathbf{V}_1$ and $\mathbf{V}_2 \rightarrow - \mathbf{V}_2$), 
or by a permutation of the two particles ($1 \leftrightarrow 2$ and $R \leftrightarrow R'$).

\section*{Appendix C. Formulas for the radial velocities}

Since the scalar products from Appendix B are Lorentz invariant, we have
\begin{eqnarray}
- R c = - \gamma'_1 \vec{v'_1} \cdot \vec{R'} - \gamma'_1 R' c, \label{Lorentzinv1} \\
\gamma_2 \, \vec{v_2} \cdot \vec{R} - \gamma_2 \, R c = - R' c, \label{Lorentzinv2} \\
\gamma_2 = \gamma'_1. \label{Lorentzinv3}
\end{eqnarray}

From equation (\ref{Lorentzinv3}) we see that $v_2^{\ 2} = {v'_1}^{\, 2} \equiv v^2$. In the particular case when
$\vec{v'_1} = - \vec{v_2}$,
the two inertial reference frames $K$ and $K'$ are linked by a Lorentz boost and its inverse transformation.

From equation (\ref{Lorentzinv2}) we get the radial component of the velocity of particle 2, as seen in reference frame $K$.
\begin{equation}
v_{2 rad} = \frac{\vec{v_2} \cdot \vec{R}}{R} = \Big(1 - \frac{R'}{\gamma_2 \, R}\Big) c. \label{eq:C63}
\end{equation}

From equation (\ref{Lorentzinv1}) we get the radial component of the velocity of particle 1, as seen in reference frame $K'$.
\begin{equation}
v'_{1 rad} = \frac{\vec{v'_1} \cdot \vec{R'}}{R'} = \Big(- 1 + \frac{R}{\gamma'_1 \, R'}\Big) c. \label{eq:C64}
\end{equation}

From equation (\ref{eq:C63}) we write
\begin{equation}
\frac{R'}{R} = \Big(1 - \frac{v_{2 rad}}{c}\Big) \gamma,
\label{eq:C65}
\end{equation}
and from equation (\ref{eq:C64}) we write
\begin{equation}
\frac{R}{R'} = \Big(1 + \frac{v'_{1 rad}}{c}\Big) \gamma,
\label{eq:C66}
\end{equation}
where $\gamma = 1 / \sqrt{1 - v^2 / c^2}$
is the common value seen in equation (\ref{Lorentzinv3}).
Due to equation  (\ref{eq:ratio2}), we see that 
equations (\ref{eq:C65})-(\ref{eq:C66}) are consistent with equations (\ref{eq:galeriu1})-(\ref{eq:galeriu2}).
By multiplying the above two equations we obtain
\begin{equation}
\Big(1 - \frac{v_{2 rad}}{c}\Big) \Big(1 + \frac{v'_{1 rad}}{c}\Big) = \frac{1}{\gamma^2} = 1 - \frac{v^2}{c^2}.
\label{eq:rad_vel_prod}
\end{equation}

Is it possible for the two radial velocities $v_{2 rad}$ and $v'_{1 rad}$ to be equal? 
If $v_{2 rad} = v'_{1 rad} \equiv v_{rad}$, then from equation (\ref{eq:rad_vel_prod}) it follows that $v_{rad}^2 = v^2$,
which means that each particle, relative to the other particle, must move only in the radial direction.

\end{document}